\documentclass[
 aip,
jap,
numerical,
 amsmath,amssymb,
reprint,%
floatfix,
nofootinbib,
]{revtex4-1}

\usepackage{amssymb} 
\usepackage{amsthm}
\usepackage{dcolumn}
\usepackage{bm}
\usepackage{cancel}
\usepackage{float}
\usepackage{graphicx,hyperref} \hypersetup{colorlinks=true, linkcolor=blue, urlcolor=blue, citecolor=blue}
\usepackage{url}
\usepackage{xcolor}
\usepackage{epstopdf}
\usepackage{lineno}

\begin{document}

\preprint{AIP/123-QED}

\title{miniTimeCube as a neutron scatter camera}

\author{Glenn R. Jocher}
\email{glenn.jocher@ultralytics.com}
\affiliation{Ultralytics LLC, Arlington, VA 22203, USA}

\author{John Koblanski}
\email{johnk2@hawaii.edu}
\affiliation{Department of Physics and Astronomy, University of Hawaii, Honolulu, HI 96822, USA}

\author{Viacheslav A. Li}
\email{vli2@hawaii.edu}
\affiliation{Lawrence Livermore National Laboratory, Livermore, CA 94550, USA}

\author{Sergey Negrashov}
\affiliation{Department of Physics and Astronomy, University of Hawaii, Honolulu, HI 96822, USA}
\affiliation{Department of Information and Computer Sciences, University of Hawaii, Honolulu, HI 96822, USA}

\author{Ryan C. Dorrill}
\affiliation{Department of Physics and Astronomy, University of Hawaii, Honolulu, HI 96822, USA}

\author{Kurtis Nishimura}
\affiliation{Department of Physics and Astronomy, University of Hawaii, Honolulu, HI 96822, USA}

\author{Michinari Sakai}
\affiliation{University of California, Berkeley, CA 94720-7300, USA}
\affiliation{Lawrence Berkeley National Laboratory, Berkeley, CA 94720-8153, USA}

\author{John G. Learned}
\affiliation{Department of Physics and Astronomy, University of Hawaii, Honolulu, HI 96822, USA}

\author{Shawn Usman}
\affiliation{Department of Geography and Geoinformation Science, George Mason University, Fairfax, VA 22030}

\date{\today}

\begin{abstract}

We present Monte Carlo (MC) simulation results from a study of a compact plastic-scintillator detector suitable for imaging fast neutrons in the 1 -- 10 MeV energy range: the miniTimeCube (mTC). Originally designed for antineutrino detection, the mTC consists of 24 MultiChannel Plate (MCP) photodetectors surrounding a 13 cm cube of boron-doped plastic scintillator. Our simulation results show that waveform digitization of 1536 optically sensitive channels surrounding the scintillator should allow for spatiotemporal determination of individual neutron-proton scatters in the detector volume to $\thicksim$ 100 picoseconds and $\thicksim$5 mm. A Bayesian estimation framework is presented for multiple-scatter reconstruction, and is used to estimate the incoming direction and energy of simulated individual neutrons. Finally, we show how populations of reconstructed neutrons can be used to estimate the direction and energy spectrum of nearby simulated neutron sources.

\end{abstract}

\keywords{simulation, estimation, optimization, fitting, neutron, gamma, detection, direction, energy, scintillation}
\maketitle

\tableofcontents

\section{\label{sec:level1}Introduction}

Recent advances in methods to reconstruct particle interactions in scintillating media\cite{jgl09} coupled with advanced picosecond (ps) resolution photodetectors and electronics \cite{ber13} will allow for new instruments with unprecedented four-dimensional (x, y, z, t) imaging of antineutrinos and neutrons emanating from nuclear reactors and radionuclides, respectively. For antineutrinos this results in improved spatiotemporal resolution of the positron and neutron associated with the Inverse Beta Decay (IBD) reaction. While for neutrons, it allows for observation of  ``light starved'' proton-proton scatters. Such a methodology allows for the reconstruction of particle kinematics (vertex and energy deposition) and subsequent estimation of the particle topology and directionality. This capability has significant scientific applications, in such areas as the study of neutrino oscillations phenomena and the search for sterile neutrinos.

The primary purpose of this paper is to address the issue of fast neutron directional detection in plastic scintillator instrumented with fast electronics and high photocoverage. However, in this introduction we also provide information of why it is important to build multi-purpose detectors, which can be used to monitor nuclear reactor via antineutrino inverse-beta-decay detection and to detect special nuclear materials via fast-neutron re-scattering.

\subsection{Reactor antineutrinos}

There are significant practical applications of such developments, primarily in the imaging of antineutrinos to geolocate nuclear reactors and detecting neutron scatters to find Special Nuclear Materials (SNR) from increased stand-off distances. Additionally, as part of the development program, study of spatiotemporal resolution will improve algorithmic discrimination of cosmogenic and geophysical backgrounds (which usually plague prior types of neutrino and neutron detectors). 

Thus, we present a description of the miniTimeCube (mTC), a proof of concept antineutrino and neutron imager. The detector consists of a $(13\ \mathrm{cm})^3$ Boron-doped plastic cube surrounded with four detectors (Photonis 64 pixel multichannel plate photomultipliers) per face, operating in the 850 ps rise-time range ($\thicksim60$ picoseconds single-PhotoElectron (PE) time resolution with waveform fitting). The detector was demonstrated at the NIST Center for Neutron Research (NCNR) in Gaithersburg, Maryland. 

For antineutrinos, the majority of reactor based neutrino experiments have traditionally employed the use of large scale water/scintillator based detectors. The concept of using small neutrino detectors ($<5$~kt) near nuclear reactors to monitor nuclear power and fuel burn up in the core was introduced in 1970s, but was technically impractical then. Subsequently, in Russia the Rovno neutrino spectrometer for the first time observed the distortion of the antineutrino spectrum as a function of uranium burnup and plutonium accumulation within the core \cite{kli94}. 

The KamLAND kiloton scintillation detector started operation in 2002 and recorded neutrinos from reactors all around Japan, definitively determining the oscillations of electron anti-neutrinos.  The similarly-sized Canadian experiment SNO simultaneously settled the solar neutrino anomaly (missing neutrinos) in favor of electron neutrino oscillations; and the smaller Borexino experiment in Italy clinched the issue. In more recent years detectors at 100~m to 2~km from reactors, as at Double CHOOZ, Reno and Daya Bay have been employed to constrain oscillation parameters at various baselines \cite{dchoo,reno,daya}, completing our first order understanding of three-neutrino mixing.

The San Onofre Nuclear Generating Station detector (SONGS1) collaboration proposed \cite{ber02} and demonstrated \cite{bow07} a compact (about one tonne) antineutrino detector specifically designed for International Atomic Energy Agency (IAEA) nuclear safeguards applications.  They demonstrated the possibility of monitoring a (cooperating) nuclear reactor in neutrino output from tens of meter ranges, employing fairly simple modern technology.  Note that the special virtue of such observations is that such are completely non-invasive, and that neutrinos cannot be faked or hidden. Several types of detectors have been tested at the now abandoned SONGs station but only one yielded useful results, emphasizing that such development is not completely trivial even with modern equipment.

Recently, a reanalysis of all reactor neutrino experiments since the original Reines and Cowan discovery of neutrinos in the 1950's, has shown a $\thicksim6\%$ deficit in the expected antineutrino flux at baselines $<100$~m compared to calculated fluxes, which is referred to as the Reactor Antineutrino Anomaly (RAA) \cite{men11}. The RAA coupled with a renewed interest in counter-proliferation have generated an increased focus on small-scale neutrino detectors.

Several small-scale detectors have been built to address non-proliferation, short baseline oscillations, and the recent evidence of the reactor antineutrino anomaly. Given the compact size of the detectors, they are typically deployed near the reactor core, typically within 100 meters. Experiments at such short baselines experience extremely high reactor-induced backgrounds that are otherwise negligible at longer baselines. 

For long distances, there is another experiment, currently under construction --- WATer CHerenkov Monitor of ANtineutrinos (WATCHMAN~\cite{Askins:2015bmb}). It will be the first prototype detector to remotely monitor individual reactor operations at a significant distance.

The three main backgrounds of concern are gamma radiation, fast neutrons, and cosmogenic muons. Lead or other heavy metal, and polyurethane or wax shielding is often used to mitigate reactor-induced gamma and neutron radiation. In addition to passive shielding, muon paddles (large flat scintillation counters) are typically used to tag muons.  Muons may pose a problem for electronic dead time for larger detectors ($> 1$~m$^3$) which are not deeply buried. But muons are also problematic to antineutrino detection even in deep detectors due to the occasional production of long lived isotopes whose decay a second later can fake a neutrino interaction.

\subsection{Neutron detection}

For neutron imaging, eleven orders of magnitude separate the IBD cross section and the neutron-proton cross section mechanism. Thus, indefinite volumetric expansion of neutron detectors is constrained due to the diminishing returns in rate measurements, since the neutrons will not deeply penetrate the detector volume. The first directional neutron scatter camera measured the double proton recoil in two separate planes of plastic scintillator to determine a crude estimate of the angular resolution of the neutron \cite{van07}. Using this method, first presented in 1985 \cite{pre85,wal96}, neutron energy can be determined through a combination of peak amplitude measured in the primary plane followed by a time-of-flight measurement in the second plane. Leveraging the estimated locations of proton scatters in each plane in conjunction with the energy lost from the pulse amplitude in the 1st plane produces a cone of directionality whose angle is determined by rudimentary kinematic equations for elastic scattering. Similarly, a neutron scatter camera using two spatially separated liquid scintillator cells demonstrated the capability of detecting a $^{252}$Cf source at a 30m stand-off distance \cite{mas09}.  Pinhole apertures and an array of coded pinhole apertures have also successfully imaged fissile neutrons \cite{bla11,bru09}. However, the pinhole aperture approach to directionality are significantly limited due to the low count rates associated with achieving directionality through restriction of the observing solid angle area of the aperture.  Our approach differs from the preceding primarily because we use a single volume to record both scatters.

Similar to cloud chambers or bubble chambers, Time Projection Chambers (TPCs) employ a detector volume filled with a gas or liquid or combination of gases allow for the high resolution imaging of particle scatters. Bubble and cloud chambers permit the photography of the particle induced disturbances. TPCs employ an electric field to drift the ions and electrons to electronic pickups, making a sort of time sequence of the image projected onto a plane. Our concept with the mTC aims at using the now available high precision light detection time to conceptually combine these methods, in a sort of photon-TPC.  Key to this method, now being explored for use in higher energy detectors as well as here at nuclear energies, is the realization of Fermat's Principle, which states that light will always take the shortest path from object to image.  In other words, despite the relatively long emission times of scintillating materials, the leading photons will reveal the event topology.  In the past we have mainly thought of scintillation detectors as only useful for calorimetry, measuring the total light output and thus energy since the scintillation radiation is isotropic. It has been demonstrated that fast light detection can pinpoint interaction locations (the vertex) and even permit track reconstruction and particle identification~\cite{Barrett:2013zka}.

As applied to neutron detection, such a methodology is advantageous compared to the aforementioned neutron cameras because it allows for the three-dimensional reconstruction and ionization of the recoiling proton. Helium-3 has typically been considered the ``gold standard" for neutron detection and TPCs in general due to the high neutron detection efficiency, gamma radiation discrimination \cite{kou09}, non-toxicity, and low cost \cite{gao11}. However, the recent shortage in global helium supplies \cite{she10} has resulted in an increased level of research activity to innovate neutron detectors to mitigate the lack of $^3$He. Several successful $^2$H TPC alternatives \cite{smi05, bow10, hun08} have been demonstrated however they have not been widely accepted into the safeguards community due to safety concerns. Gas filled TPCs provide an excellent means of imaging fast neutrons, however they suffer from a low detector cross section due to the $\sim10^{-3}$ decreased density of gas as compared to plastic. Such detectors also do not promise to allow for robust operation in a field deployment (wire vibration, high voltage, ultra-cleanliness, and high pressure special gases and containers are problematic).

The mTC aims to overcome these shortcomings. Key to this is also the availability of relatively low cost, high density and very fast (ps) digitization electronics, developed at the University of Hawaii. While we originally developed the concept for \textit{antineutrino} imaging\cite{li2016}, the mTC could also potentially image fast neutrons from radionuclides with performance comparable to traditional gas-based TPCs. We leverage a robust GEANT4 and MATLAB Monte Carlo simulation, validated by real world data \cite{jocher2013}, which shows the mTC's ability to image fast neutrons with a high efficiency and significant angular accuracy. 

\section{Detector Description}
\label{Detector Description}

\begin{figure}[htbp!]
\includegraphics[width=\linewidth]{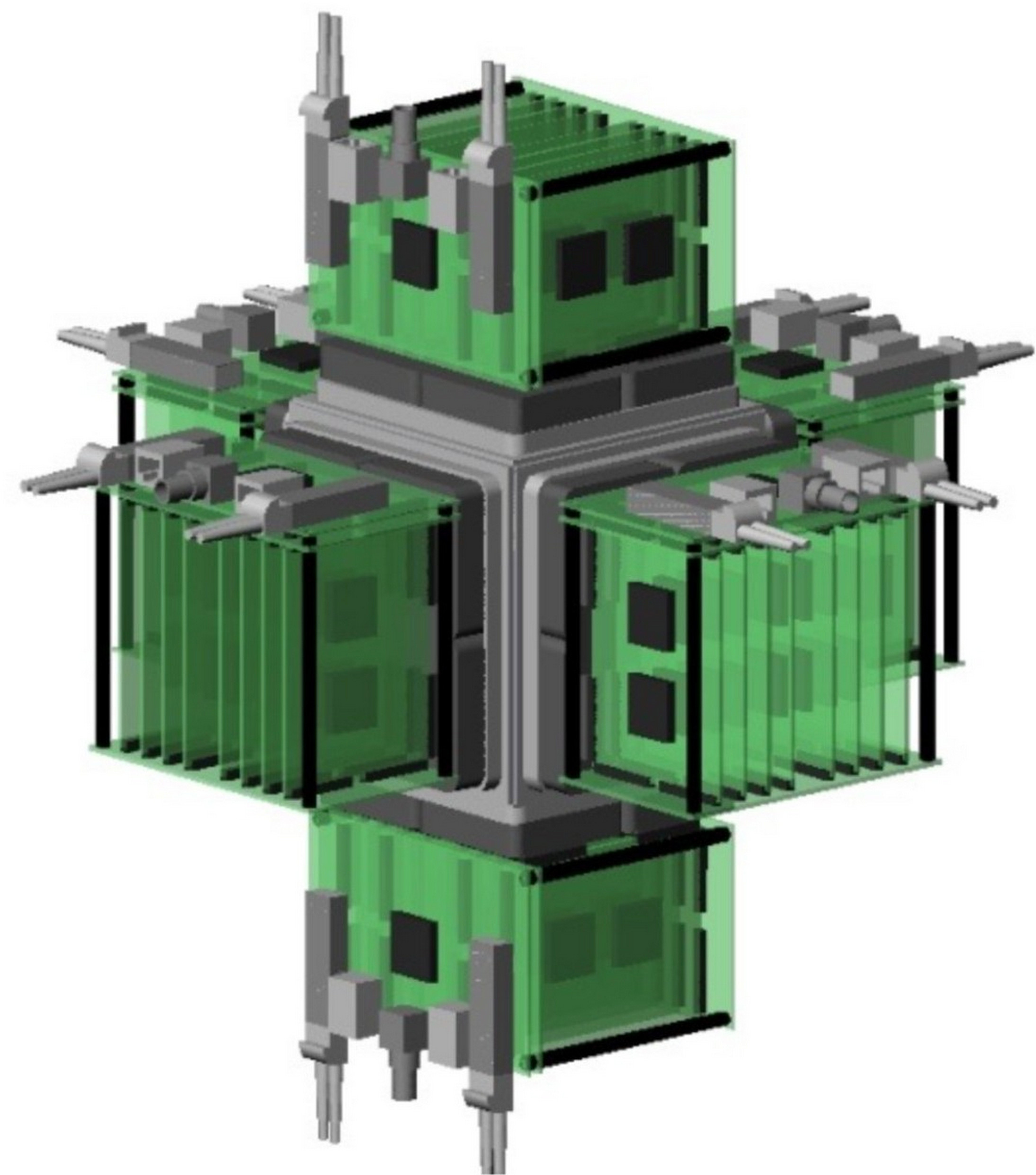}
\caption{CAD drawing of mTC showing the plastic scintillating cube enclosed on all 6 sides by MCP-PMTs and digitizing electronics.}
\label{mTC photo}
\end{figure}

The mTC is a $13 \mathrm{cm} \times 13 \mathrm{cm} \times 13 \mathrm{cm}$ cube, composed of 2.2 liters of  1\% natural boron-doped (0.2\% $^{10}$B by weight) EJ-254\cite{EJ-254} plastic scintillator surrounded by 24 Planacon Micro-Channel-Plate Photomultiplier Tubes (MCP-PMTs). Each MCP-PMT has 64 unique channels, for 256 channels per face and 1536 total across all 6 sides. Figure \ref{mTC photo} shows the mTC with surrounding MCP-PMTs and all associated electronics. Each channel is coupled to full-waveform readout electronics sampling at  $\sim$2.8 GigaSamples Per Second (GSPS), which are theoretically capable of providing $<$~100~ps timing resolution of single Photo Electrons (PEs).

During the mTC operation, we investigated different trigger schemes~\cite{Li:2018dwm}. The final version was a 2-level trigger --- with preset threshold level (L0) and minimum and maximum numbers of triggered channels (L1 min and L1 max).

\section{Neutron Imaging Theory}
\label{Neutron Imaging Theory}

When a neutron in motion is deflected from a straight line trajectory, the angle of the deflection may be calculated given the kinetic energy of the neutron prior to and immediately following the deflection. This simple relationship is expressed by Equation \ref{eqn1}:

\begin{equation}
\label{eqn1}
\sin\Theta = \sqrt{\frac{E_0-E_1}{E_0}} = \sqrt{\frac{dE_0^1}{E_0}}
\end{equation}

\noindent where $\Theta$ is the neutron deflection angle, $E_0$ is the original neutron energy and $E_1$ the neutron energy after deflection; $dE_0^1$ being the difference. $dE_0^1$ is directly observable, however $E_0$ is not, and must be inferred via Equation \ref{eqn2}:

\begin{equation}
\label{eqn2}
E_0 = dE_0^1 + E_1
\end{equation}

\noindent where the posteriori kinetic energy $E_1$ is calculate via the neutron velocity following deflection:

\begin{equation}
\label{eqn3}
E_1 = \frac{1}{2}mv^2
\end{equation}

\noindent where

\begin{equation}
\label{eqn4}
v = \frac{dx}{dt} = \frac{||P_2-P_1||}{t_2-t_1}
\end{equation}

\begin{figure}[htbp!]
\includegraphics[width=\linewidth]{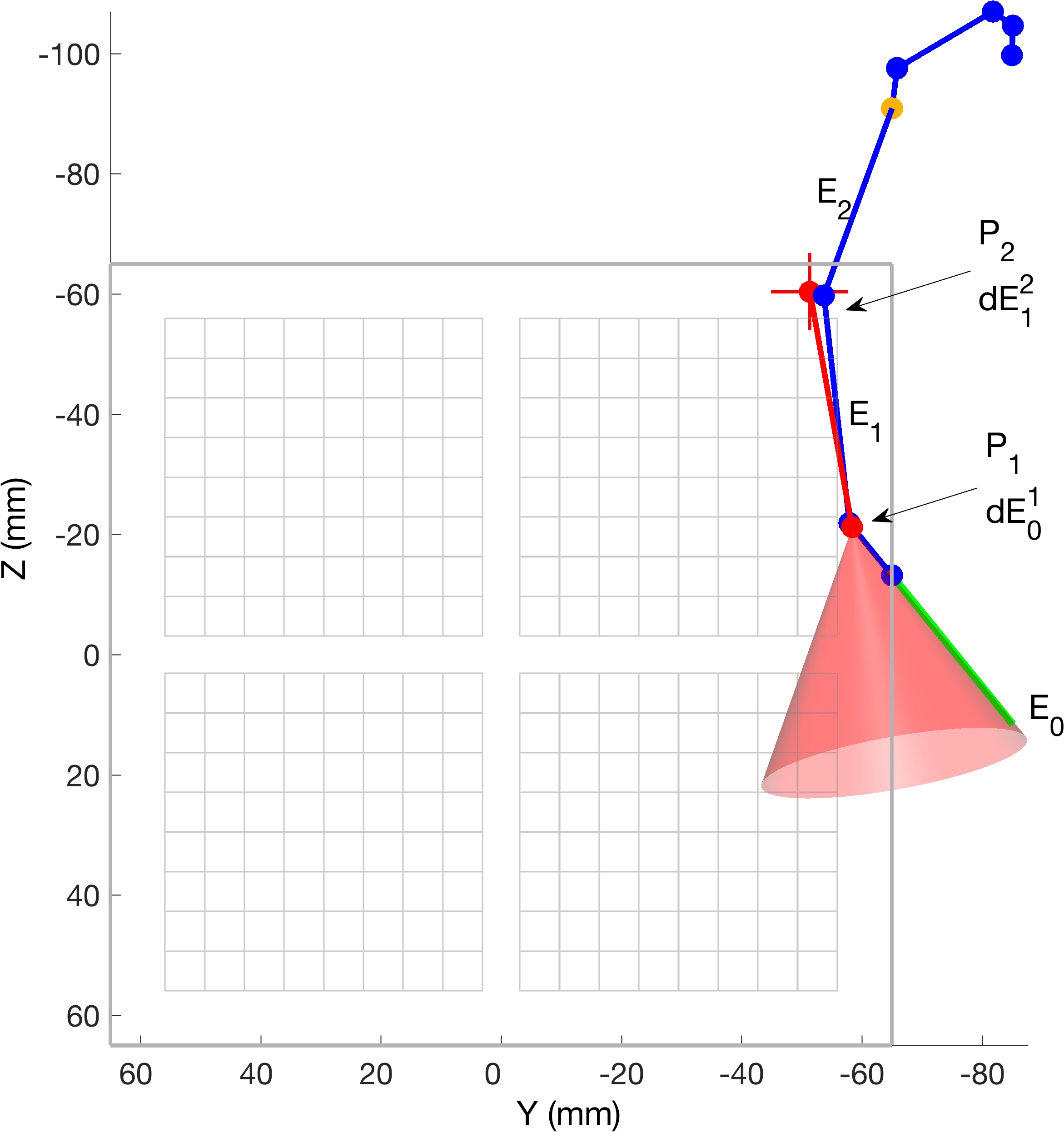}
\caption{Neutron estimation diagram. Incoming neutron in green, and true travel path through the detector in blue. Estimated travel path in red. Equation \ref{eqn1} angle cone $\Theta$ about $P_2 - P_1$ vector also shown in red.}
\label{diagram}
\end{figure}

\begin{figure}[htbp!]
\includegraphics[width=\linewidth]{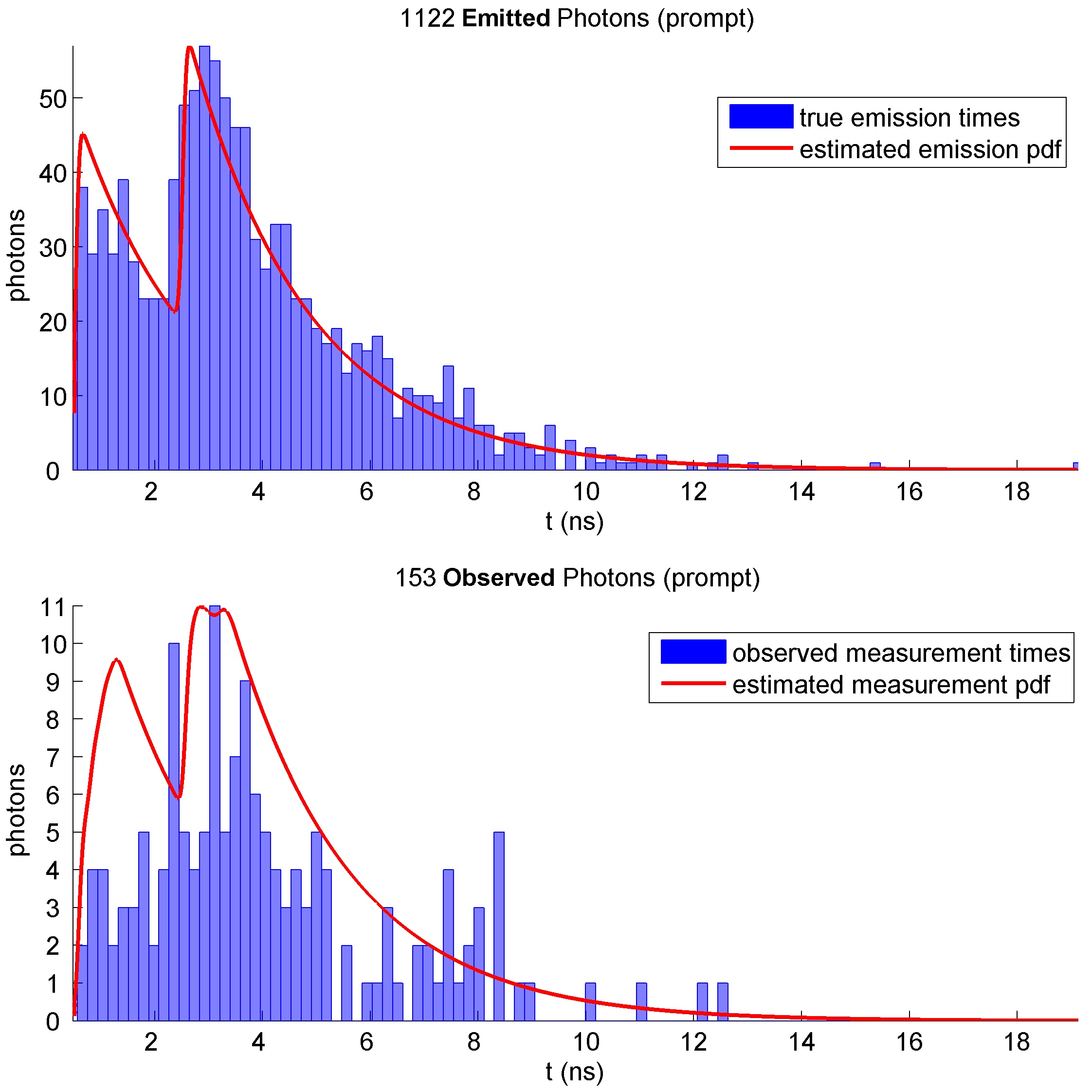}
\caption{Estimated photon emission and observation PDFs overlaid with the true distribution. The two peaks seen in this figure correspond to $P_1$ and $P_2$, occurring at times $t_1$ and $t_2$. Scintillator decay constant of 2.2 ns assumed by both truth model and estimator. Note the mixture distribution inherent in this multi-scatter estimation problem; in most cases it is impossible to completely segregate which photon (measurement) corresponds to which scatter (parameter).}
\label{measuredPhotons}
\end{figure}

To solve for $v$, one needs to know $P_1$ and $P_2$ (shown in Figure \ref{diagram}), the positions of the first and second scatters, and $t_1$ and $t_2$, the associated scatters times (visible in Figure \ref{measuredPhotons}). Neutron mass $m = 939.565378$ MeV/c$^2$ and speed of light $c = 299.792458$ mm/ns are assumed. Substituting \ref{eqn2}, \ref{eqn3} and \ref{eqn4} into \ref{eqn1} allows us to directly solve for $E_0$ and $\Theta$ from the available observables.

\section{Neutron Imaging in Practice}
\label{In Practice}

To successfully apply the neutron direction and energy method described in this paper, 3 observable parameters are required from scatter 1:

\begin{enumerate}\itemsep1pt \parskip0pt \parsep0pt
\item Position $P_1$
\item Time $t_1$
\item Neutron kinetic energy lost during scatter 1: $dE_0^1$ 
\end{enumerate}

\noindent and 2 additional parameters from scatter 2:
\begin{enumerate}\itemsep1pt \parskip0pt \parsep0pt
\item Position $P_2$
\item Time $t_2$
\end{enumerate}

In practice none of these parameters are perfectly knowable; however, we may attempt to estimate their values from the evidence the neutron leaves behind. This evidence comes in the form of photons, light emitted by the charged particles that neutrons tend to bump into in organic scintillator. Light sensors surrounding the scintillator can record the timing and location of these photon bursts, and this information can in turn be used to estimate the 5 parameters above.

Organic scintillator is composed of long hydrocarbon chains with loosely bonded protons at the perimeter. Neutrons traveling through these chains will tend to hit the loosely bonded protons about 60\% of the time, colliding with $^{12}$C atoms the rest of the time (A small minority ($\sim$1\%) of scatters occur on C$^{13}$ and other isotopes).

When a neutron hits a proton in the scintillator, it transfers some of its kinetic energy to the proton, which in turn becomes visible light via charged particle ionization of the surrounding atoms the proton passes by. The proton path is usually short, a few mm, and the scintillation yield is usually low, due to the large mass of the proton, yet a high energy ($\sim$ 1~MeV) neutron recoil may still yield significant observable light.

\section{Observing Neutrons in the mTC Detector}

The real world observability, or `resolution', of each of these 5 parameters ultimately dictates neutron $\Theta$ and $E_0$ observability, therefore we seek to estimate these values as accurately as possible for every incoming neutron. To this purpose we have created estimators and verified their performance via MC simulation of a real world detector: the UH miniTimeCube (mTC) antineutrino detector. Though intended primarily for antineutrino research, the fast timing of this detector, long recording buffer, and high neutron cross section make it well suited to neutron reconstruction as well, thus we use it as our modeled testbed for simulated neutron reconstruction.

\subsection{Individual Neutron Estimation}

To fit a single neutron, we apply 5 parallel estimators to the measurements, with each estimator assuming a different number of observable neutron scatters within the detector. The Maximum A Posteriori (MAP) likelihood and parameter count of each estimator are then used to compute the corresponding Bayesian Information Criterion (BIC), and the smallest BIC of the 5 indicates the most likely number of scatters. If 2 or more scatters are determined by the BIC, then the first two scatters are isolated and used to compute neutron $\Theta$ and $E_0$ per Equation \ref{eqn1}. If the BIC determines only 1 scatter occurred, the event is not considered a candidate and is rejected.

Each estimator fits the MAP values of the 5 parameters ($x$, $y$, $z$, $t$, $E$) which fully describe each recoil. For $n$ recoils the parameters are:

\begin{eqnarray}
\label{eqnEst1}
\theta_n = \left[ \begin{array}{ccc}P_1 & t_1 & w_1	\\
& \vdots &		\\
P_n & t_n & w_n\end{array} \right]
\end{eqnarray}

\noindent where $w$ indicates the weight of each scatter. For a finite set of scatters, the non-negative, nonzero weights $w>0$ are constrained to a convex combination summing to unity:

\begin{equation}
\label{eqnEst2}
\sum\limits_{i}^{n} w_i = 1
\end{equation}

The measurements available for this task, $z$, are the observed photon arrival positions $P^\gamma$ and times $t^\gamma$ on the MCPs surrounding the scintillator:

\begin{eqnarray}
\label{eqnEst3}
z = \left[ \begin{array}{cc}P_1^\gamma & t_1^\gamma	\\
\vdots & \vdots		\\
P_m^\gamma & t_m^\gamma \end{array} \right]
\end{eqnarray}

The measured positions are discretized to the MCP anode centers (each channel is a 5.9 mm $\times$ 5.9 mm square). Measurement time resolutions are determined through MC study, in the case of the mTC we find these to be $<$ 100 ps $1\sigma$ per channel for first arrivals including the effect of the MCP Transit Time Spread (TTS) (we note that empirical Photonis MCP differ substantially from simulated results).

The following steps are used to fit up to 5 possible neutron scatters:

\begin{enumerate} \itemsep1pt \parskip0pt \parsep0pt
\item Assume 1 scatter and estimate $\hat{\theta}_1^{MAP}$
\item Assume 2 scatters and estimate $\hat{\theta}_{1...2}^{MAP}$
\item Assume 3 scatters and estimate $\hat{\theta}_{1...3}^{MAP}$
\item Assume 4 scatters and estimate $\hat{\theta}_{1...4}^{MAP}$
\item Assume 5 scatters and estimate $\hat{\theta}_{1...5}^{MAP}$
\item Evaluate Equation \ref{eqnBIC1} BIC for each $\hat{\theta}$
\item Scatter count defined by minimum BIC
\item Assign individual photon probabilities to each scatter
\item Estimate energy of each scatter ($dE$) using `extended Poisson' Equation \ref{poisson}
\end{enumerate}

\subsection{Mixed MAP Estimator}

The MAP cost function follows Bayes' equation\cite{Bar-Shalom}:
\begin{equation}
\label{eqnEst4}
p(\theta |z)=\frac{p(z|\theta)p(\theta)}{p(z)}=\frac{1}{c}p(z|\theta)p(\theta)
\end{equation}

\noindent where the likelihood $p(z|\theta)$ forms a mixture distribution with 1 mixture component per neutron recoil. Dropping the normalizing constant $c$:

\begin{equation}
\label{eqnEst5}
p(\theta |z) = \prod_{j}^{} \sum_{i}^{} w_i p\left(z_j|\theta_i\right)p\left(\theta_i\right)
\end{equation}

\noindent where the likelihood $p(z_j|\theta_i)$ of measurement $j$ given source $i$ with weight $w_i$ and prior $p\left(\theta_i\right)$ is simply an evaluation of the measurement space created by $\theta_i$ at $z_j$. This measurement space is defined by a point-source position $P_\theta$ at time $t_\theta$, and is a function of several detector and scintillator characteristics including:

\begin{itemize}  \itemsep1pt \parskip0pt \parsep0pt
\item Scintillation spectrum, yield and decay constant(s)
\item Cherenkov spectrum
\item Quenching factors for heavy particles
\item Scintillator attenuation length
\item Re-emission efficiency of attenuated photons
\item Refraction indices of the scintillator and PMT glass
\item PMT quantum efficiency, size and location
\item Time and energy calibrations
\end{itemize}

The likelihood of observing a single photo-electron (PE) $z_j$ at a single point due to a single point-source $\theta_i$ is

\begin{equation}
\label{eqnEst6}
p\left(z_j|\theta_i\right) = \Lambda_t P_\Omega P_\gamma P_T Q
\end{equation}

\begin{figure}[htbp!]
\includegraphics[width=\linewidth]{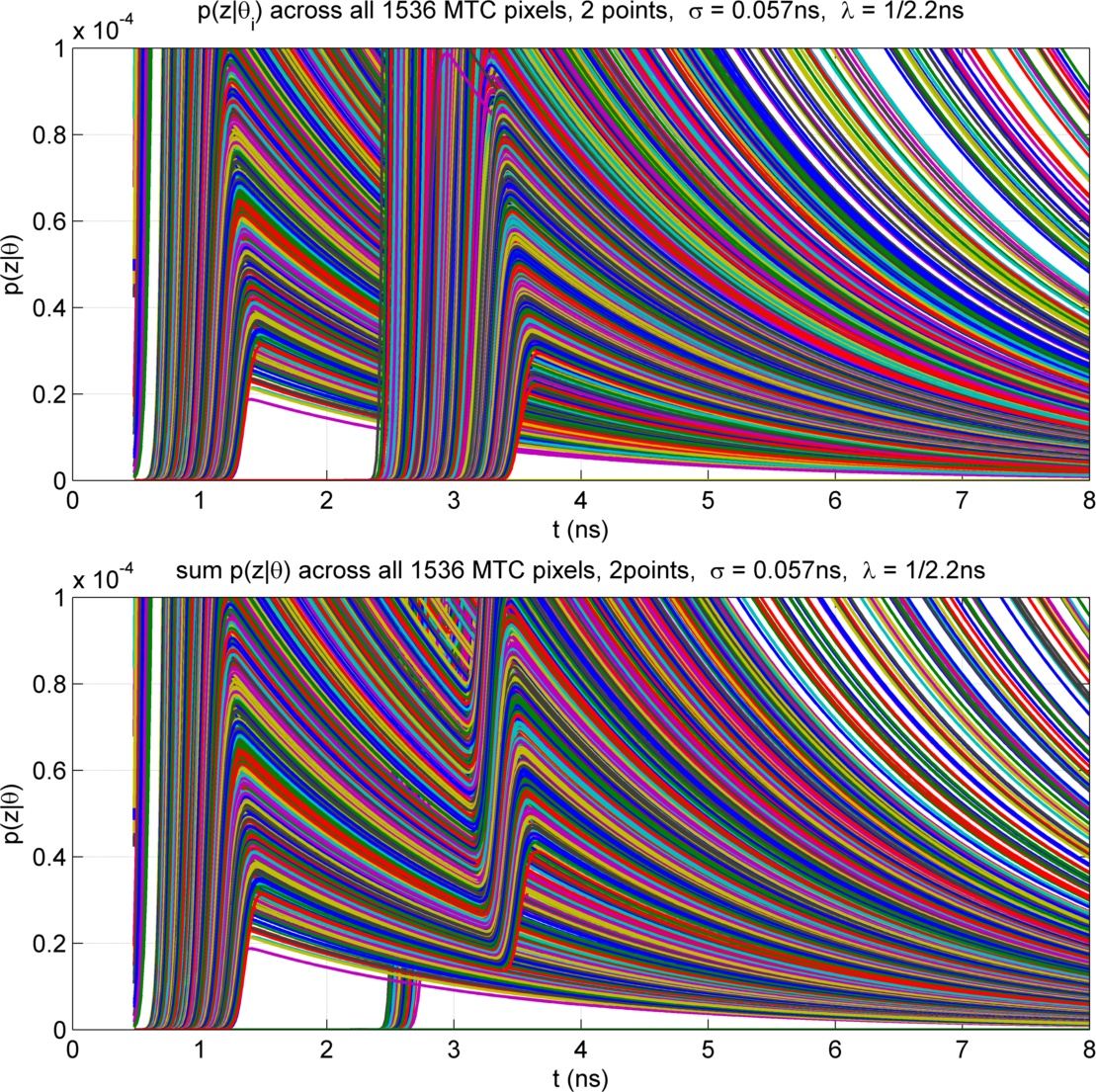}
\caption{$p(z|\theta) = \Lambda_t P_\Omega P_\gamma P_T Q$ evaluations for 2 point-sources across all channels in the mTC.}
\label{figPT}
\end{figure}

\noindent (visualized in Figure \ref{figPT}) where $\Lambda_t$ is the temporal likelihood discussed in \ref{sectionPT}, $P_\Omega$ is the solid angle probability discussed in \ref{Solid Angle}, $P_\gamma$ is the un-attenuated energy probability discussed in \ref{Attenuation}, $P_\mathrm{T}$ is the transmission probability discussed in \ref{Reflection}, and $Q$ is the PMT quantum efficiency.

\subsubsection{Scintillator Temporal Response}
\label{sectionPT}

\begin{figure}[htbp!]
\includegraphics[width=\linewidth]{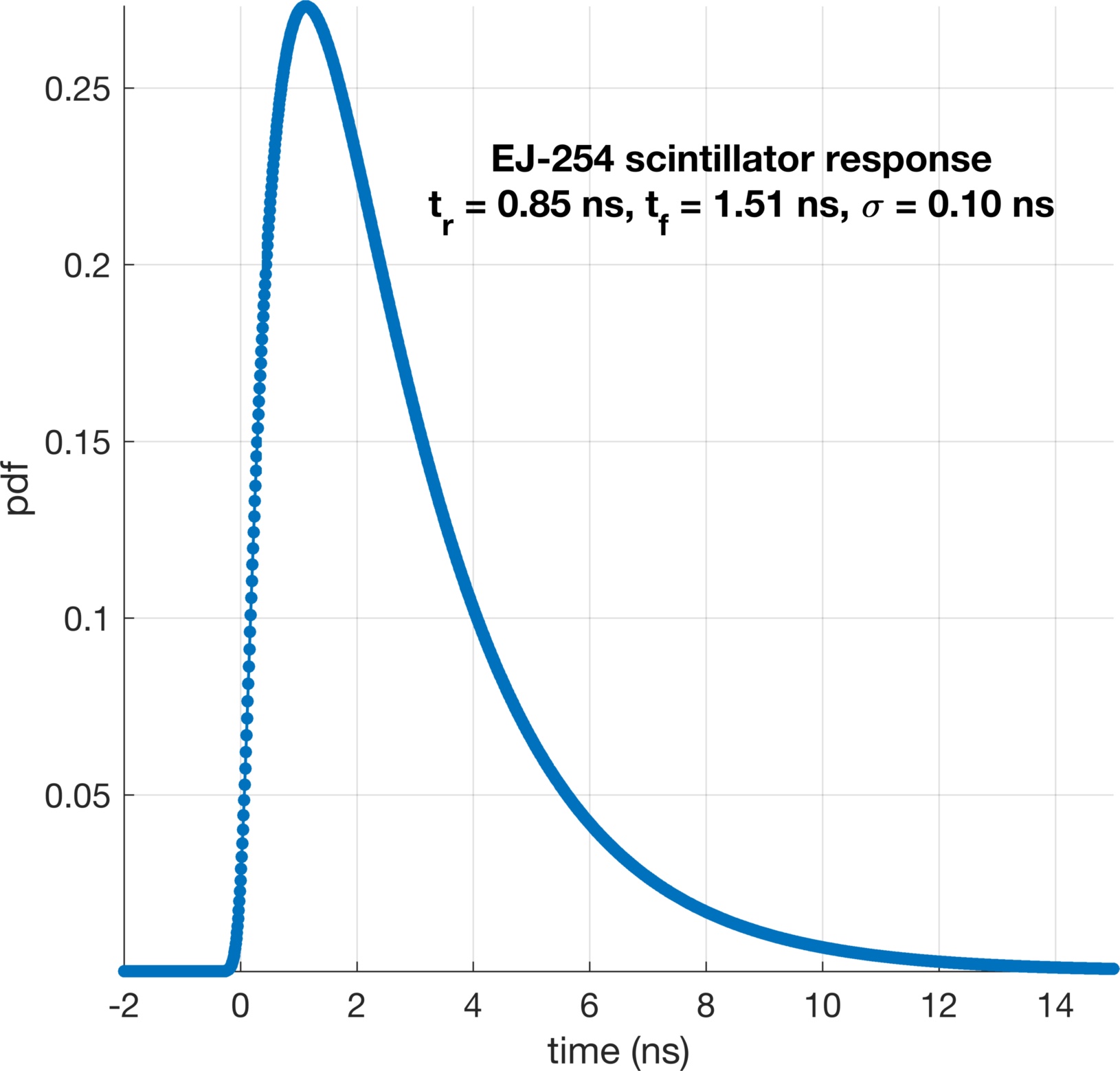}
\caption{Equation \ref{eqnPT1} evaluated over the -2 to 15 ns region. Light takes about 0.7 ns to cross the mTC's 13 cm width for comparison. A Normally distributed 0.10 ns 1$\sigma$ time smear has been applied to Equation \ref{eqnPT1} in this figure to account for PE measurement noise.}
\label{figEMG}
\end{figure}

A scintillator's temporal response is typically defined by a double exponential distribution. The probability density function (PDF) for the temporal likelihood that we use is shown in Figure \ref{figEMG}, and is:

\begin{equation}\label{eqnPT1}
\Lambda_t\left(x\right) =
\left\{
  \begin{tabular}{ll}
  $\left(1 - e^{-x/t_r} \right) e^{-x/t_f}$ & $\quad x \geq 0$ \\
  0 & $\quad x < 0$
  \end{tabular}
\right. \\
\end{equation}
where
\begin{equation}
x=t-(\theta_t+r/v)   
\end{equation}

Here $t$ is the time of the first photon arrival in channel $i$, $\theta_t$ is the estimated time of interaction, r is the distance to the MCP channel $i$, v is the velocity of light in the medium, and $t_f$ and $t_r$ are the scintillator fall time and rise time respectively.  The function acts as a prediction of measurement times for photons that propagate from a scintillation vertex. 

In the mTC's EJ-254 scintillator these times are:
\begin{equation}
  \begin{tabular}{l}
  $t_r$ = 0.85 ns \\
  $t_f$ = 1.51 ns
  \end{tabular}
\end{equation}




When used in equation \ref{eqnEst6}, the $x$ value in equation \ref{eqnPT1} is the time difference between the observation $z_j$ and the point-source $\theta_i$, after compensating for the travel time between the two points. The travel time compensation assumes a scintillation and QE weighted mean photon group speed. This will be the average speed of all {\it observed} scintillation PEs. In mTC this speed is 185.8 mm/ns.



\subsubsection{Solid Angle}
\label{Solid Angle}

\begin{figure}[htbp!]
\includegraphics[width=\linewidth]{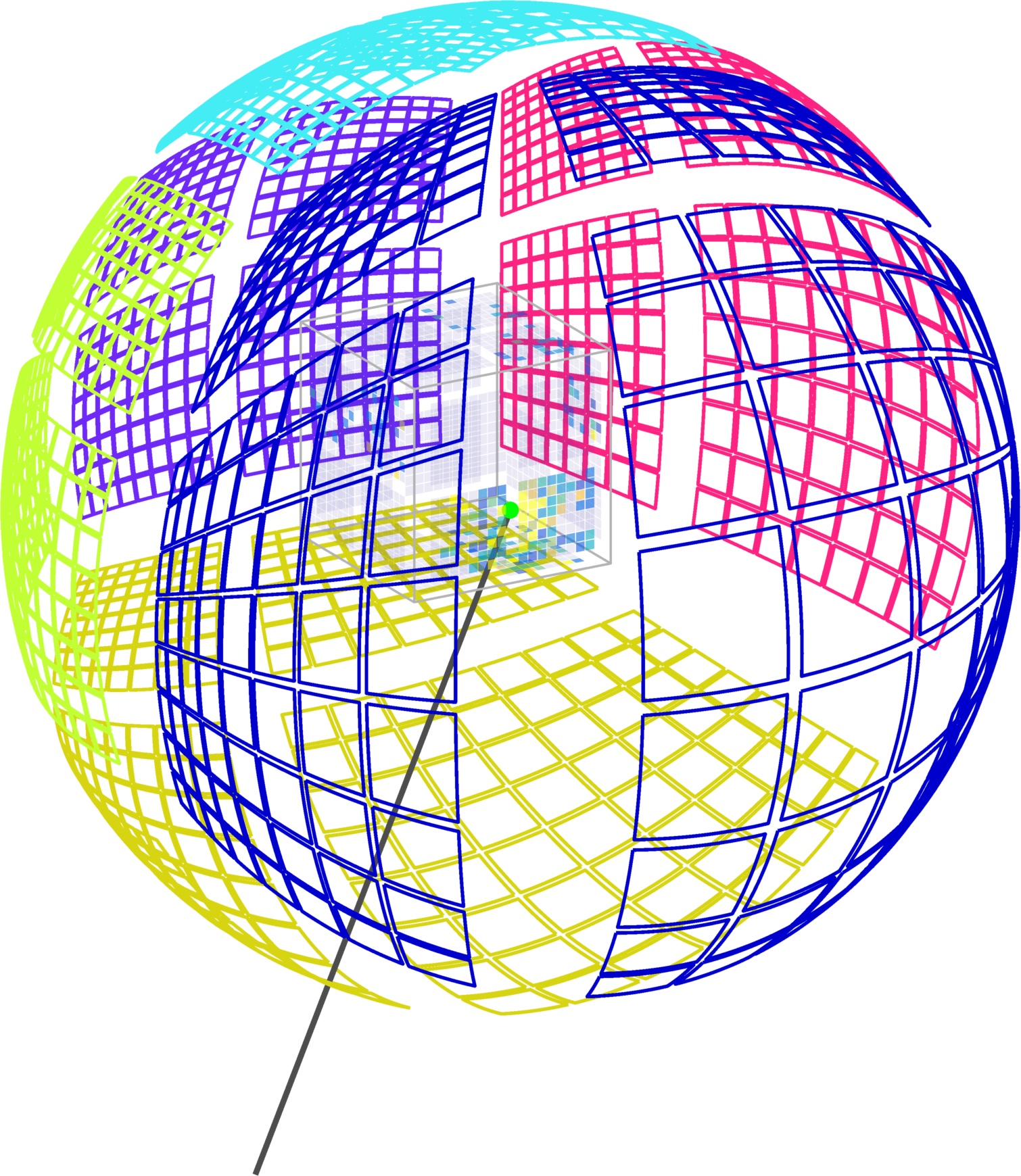}
\caption{Solid angle of all 1536 MCP channels as seen from an off-center point-source inside the mTC. The solid angle subtended by each MCP channel shown here is defined in Equation \ref{eqnSA2}.}
\label{solidangle}
\end{figure}

The solid angle of a cone with half-angle $\theta$ is 
\begin{equation}
\label{eqnSA1}
\Omega = 2 \pi \left(1 - \cos \theta \right)
\end{equation}

\noindent from which we can approximate the solid angle of an arbitrarily shaped surface $S$ (the MCP channel centered on $P_z$ in our case) with area $A$ as viewed from point $P_\theta$ by

\begin{equation}
\label{eqnSA2}
 P_{\Omega}=2\pi\left(1-\frac{r}{\sqrt{r^2+a^2}}\right)\hat{r}\cdot{\hat{n}}
\end{equation}
where $\vec{r}=\theta_{\Omega}-P_z$ is the vector pointing from $\theta_{\Omega}$ to PMT location $P_z$, $\hat{n}$ is the unit normal vector on the surface at the point $P_z$, $a$ is the radius of the surface area subtended by the cone, and $r$ is the norm of $\vec{r}$. If a certain PMT (or a group of channels) on the mTC wall observes a large percentage of the emitted photons, maximizing the solid angle probability will involve moving our point-source estimate closer to that area. The approximation comes from substituting the area of a square pixel for the radius-squared of a circular pixel. 

\begin{figure}[htbp!]
\includegraphics[width=\linewidth]{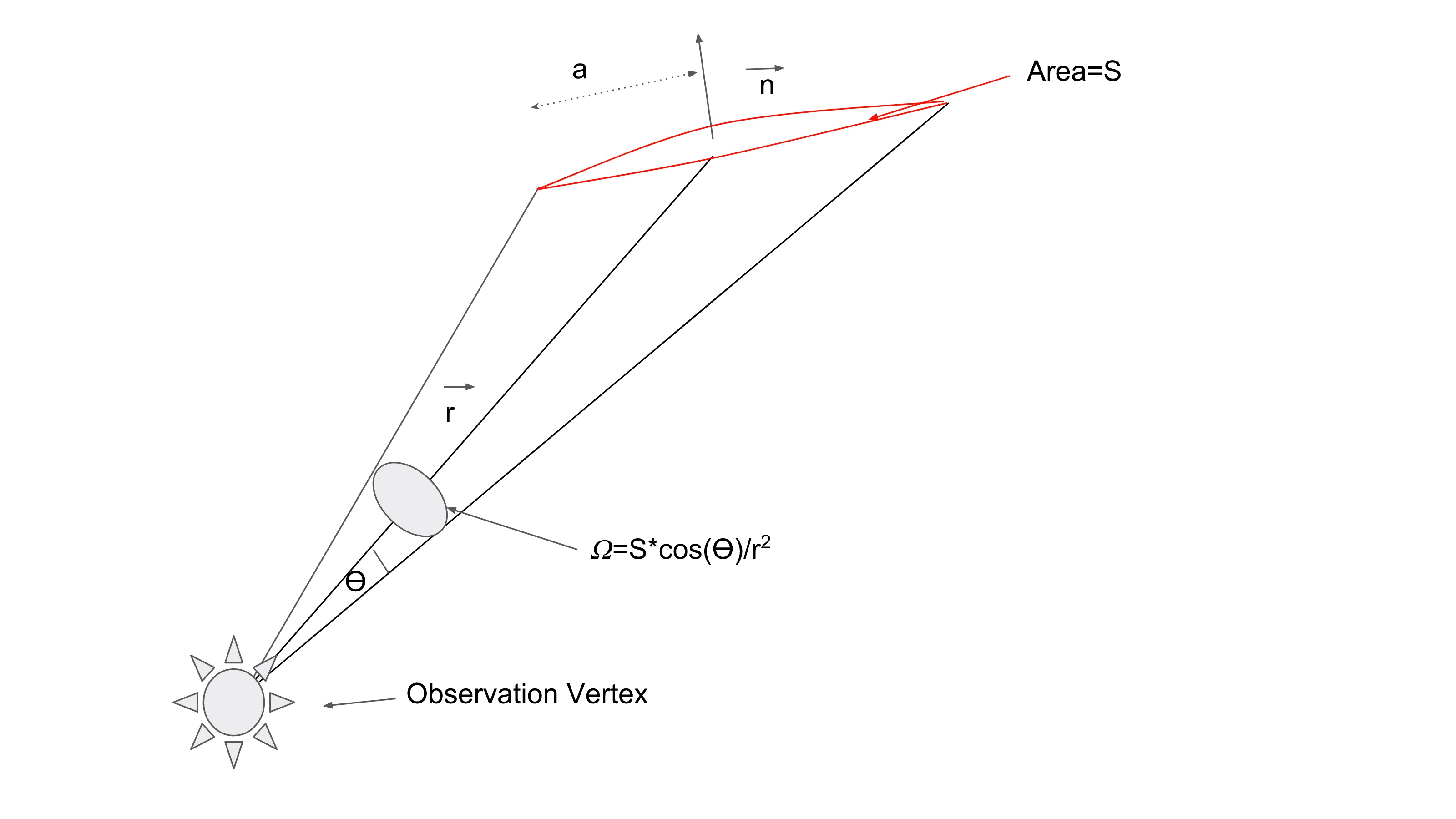}
\caption{Illustration of the solid angle, the 3d angle enclosed by a conical surface from a vertex.}
\label{fig:solid}
\end{figure}

An illustration of the mTC's solid angles from all pixels is shown in Figure \ref{solidangle}, and in general for a single source in Figure \ref{fig:solid}. 

\subsubsection{Attenuation}
\label{Attenuation}

The fraction of energy attenuation at distance $x$ from a source is given by the exponential Cumulative Distribution Function (CDF)

\begin{equation}
f\left(x;\lambda\right) = 1 - e^{-\lambda x}
\end{equation}

\noindent where $\lambda$ equals the inverse of the attenuation length of the medium being traversed. Conversely, the survival fraction is

\begin{equation}
P_\lambda = 1 - f\left(x;\lambda\right)  = e^{-\lambda x}
\end{equation}

\subsubsection{Reflection}
\label{Reflection}

When a ray of light travels between media with different indices of refraction $n_1$ and $n_2$, some incident energy will transmit through to the second media while the rest will reflect back into the first. This ratio of transmission and reflection is governed by the Fresnel equations, discovered in the early 1800's by Augustin-Jean Fresnel, a French engineer and physicist. The law of reflection holds that the reflection angle $\theta_r$ will be equal to the incident angle $\theta_i$

\begin{equation}
\theta_i = \theta_r
\end{equation}

\noindent and Snell's law defines the transmission angle $\theta_t$ of the refracted light as

\begin{equation}
\theta_t = \arcsin \left(\frac{n_1}{n_2} \sin\theta_i \right)
\end{equation}

\begin{figure}[htbp!]
\includegraphics[width=\linewidth]{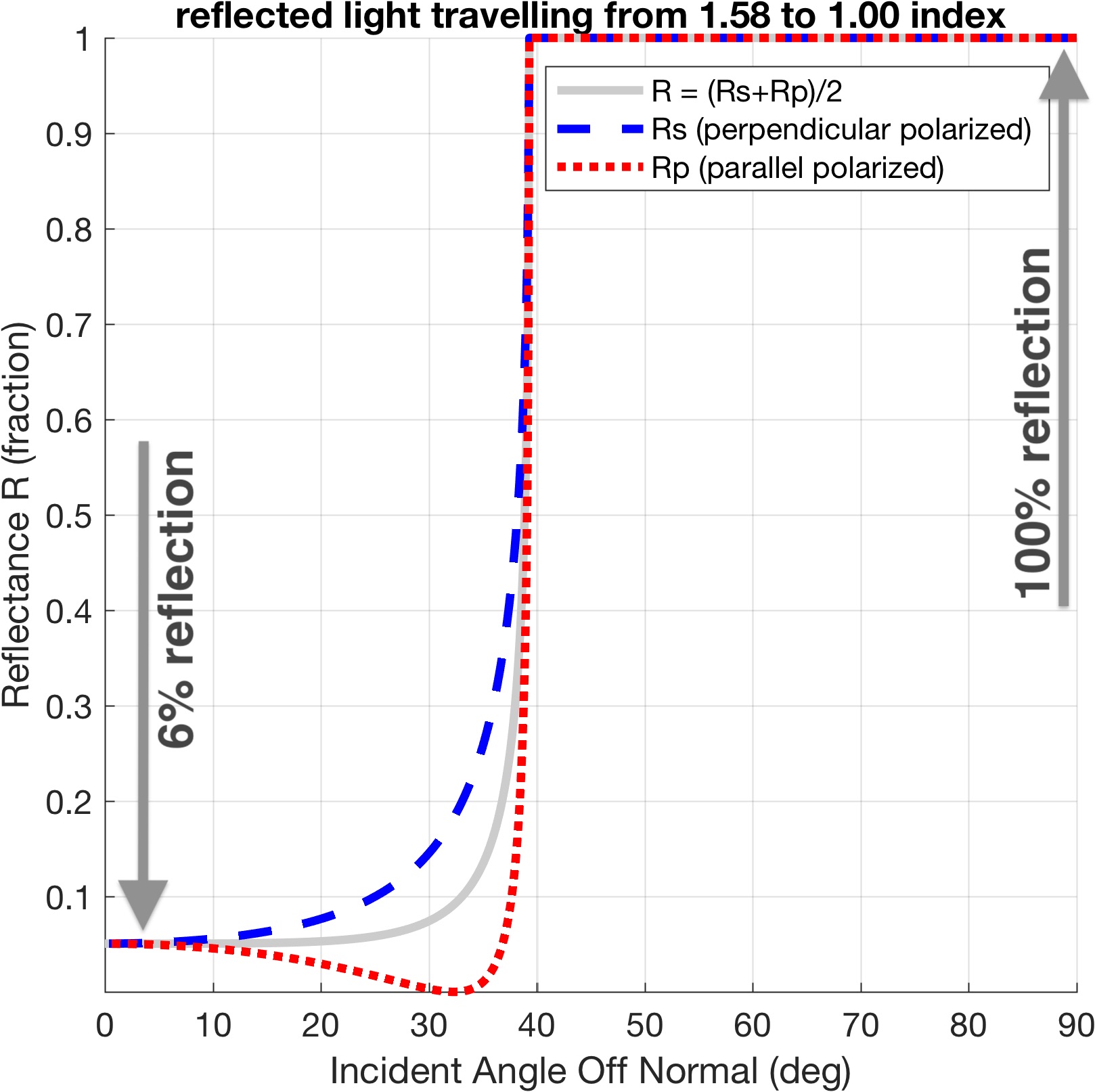}
\caption{Example reflection coefficients over incident angles 0 to $\pi/2$. }
\label{figreflection}
\end{figure}

For $s$ polarized light (incident light is polarized with its electric field perpendicular to the incident plane. The light is said to be s-polarized, from the German senkrecht (perpendicular)), the reflection coefficient (which gives the fraction of reflected light), is given by

\begin{eqnarray}
\label{eqnreflection1}
R_s &=& \left| \frac{n_1 \cos\theta_i - n_2 \cos\theta_t}{n_1 \cos\theta_i + n_2 \cos\theta_t} \right|^2	\\
&=& \left| \frac{n_1 \cos\theta_i - n_2\sqrt{1-\left(\frac{n_1}{n_2} \sin\theta_i \right)^2} }{n_1 \cos\theta_i + n_2\sqrt{1-\left(\frac{n_1}{n_2} \sin\theta_i \right)^2} } \right|^2
\end{eqnarray}

\noindent and for $p$ polarized light (incident light is polarized with its electric field parallel to the plane containing the incident, reflected, and refracted rays. Such light is described as p-polarized) the reflection coefficient is

\begin{eqnarray}
\label{eqnreflection2}
R_p &=& \left| \frac{n_1 \cos\theta_t - n_2 \cos\theta_i}{n_1 \cos\theta_t + n_2 \cos\theta_i} \right|^2	\\
&=& \left| \frac{n_1\sqrt{1-\left(\frac{n_1}{n_2} \sin\theta_i \right)^2}  - n_2 \cos\theta_i}{n_1\sqrt{1-\left(\frac{n_1}{n_2} \sin\theta_i \right)^2}  + n_1 \cos\theta_i} \right|^2
\end{eqnarray}

\noindent If the incoming light is unpolarized (with equal parts $s$ and $p$ polarization), then the reflection coefficient, shown in Figure \ref{figreflection}, is

\begin{equation}
\label{eqnreflection3}
R = \frac{R_s + R_p}{2}
\end{equation}

Via conservation of energy the transmission fraction is

\begin{equation}
T = 1 - R
\end{equation}

In cubical scintillation volumes like the mTC, reflected light can make up a significant part of the overall light collected. The overall light distribution is the sum of the original source plus the reflection source-points. To simulate $r$ reflections, $p$ reflected origins are required, as given by table \ref{table:reflections}.

\begin{table}[htbp!]
\fontsize{10}{10}\selectfont
\begin{center}
\begin{tabular}{c|c|c}
\bf Reflections & \bf New Points 	& \bf Total Points  \\
$r$             & $p = 4r^2 + 2$ 	& $n=\sum\limits_{i}^{r}p_i$ \\
\hline  
0				& 1					& 1 	  	\\
1				& 6					& 7 	    \\
2				& 18				& 25 	    \\
3				& 38				& 63 	    \\
4				& 66				& 129 	    \\
5				& 102				& 231 	    \\
\end{tabular}
\caption{$r$ reflections in a rectangular volume can be modeled by $n$ virtual points outside the detector.}
\label{table:reflections}
\end{center}
\end{table}






\subsection{Poisson Energy Estimator}

The Poisson Energy Estimator is akin to the traditional photon-counting method, the main difference being that the estimator is capable of assuming multiple point sources spread out in space and assigning expected photon counts to each pixel based on these multiple point sources.

The Poisson energy estimator estimates point-source energies conditional on a given point-source location. It can operate on a single source at a time, or estimate the combined energies of all sources jointly, which may then be apportioned to each source using its likelihood ratio. The estimator maximizes the likelihood of a product of Poisson probabilities, one for each pixel in the detector:

\begin{equation}
p(z|\theta) = \prod\limits_{i}^{} f(k_i; \lambda_i)
\end{equation}

The Poisson probability mass function (PMF) is

\begin{equation}
f(k;\lambda) = \frac{\lambda^k e^{-\lambda}}{k!}
\label{poissonEqn}
\end{equation}

\noindent where $\lambda$ is the expected number of events and $k$ the measured number. An integration through $\Lambda_t$ in Eq. \ref{eqnEst6} (along with scintillator photon yield $Y$ and PMT quantum efficiency $Q$) is sufficient to supply us with an expected number of photons at each channel over a time period

\begin{equation*} 
    \lambda_i=E*Y* p(z_i|\theta)
\end{equation*}
or,
\begin{equation} 
    \lambda_i=E*Y* P_\Omega* P_\gamma* P_T*\Lambda_t*Q
\end{equation}

\begin{figure}[htbp!]
\includegraphics[width=\linewidth]{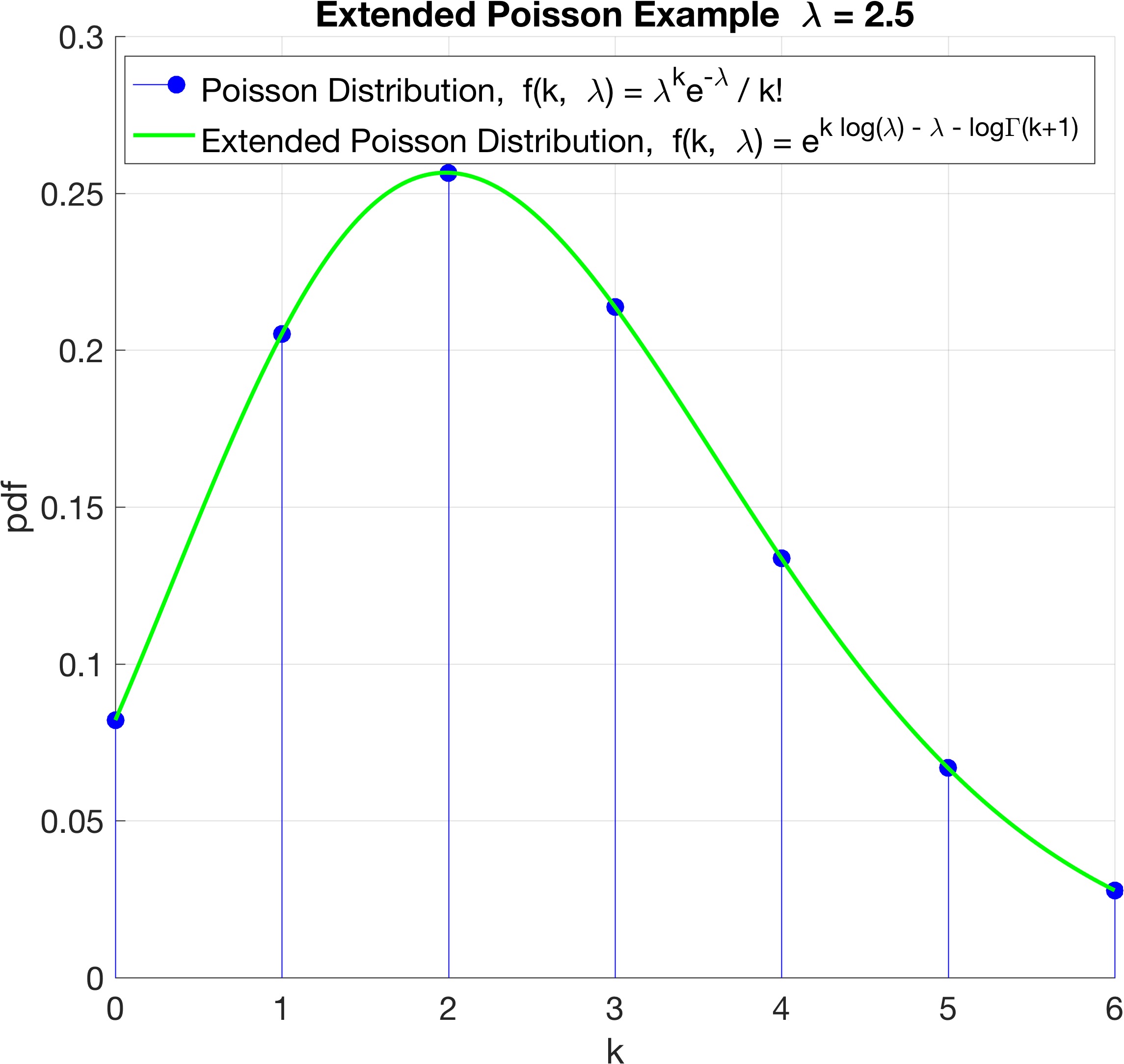}
\caption{Standard equation \ref{poissonEqn} Poisson distribution compared to the proposed equation \ref{extendedPoissonEqn} `extended Poisson' distribution.}
\label{poisson}
\end{figure}

The observed numbers of photons $k$,  are unfortunately not integers but rather fractions, as are the expected number of photons $\lambda$. A Poisson distribution is only defined for integer values of $k$, thus the problem calls for a different type of distribution capable of non-integer support. This distribution, which we dub the `extended Poisson' probability is shown in Figure \ref{poisson}, and takes on the form

\begin{equation}
f(k;\lambda) = e^{ k \log(\lambda) - \lambda - \log\left( \Gamma(k+1) \right)  }
\label{extendedPoissonEqn}
\end{equation}

\noindent where the gamma function $\Gamma$ escapes confinement by interpolating the factorial 
\begin{equation}
\Gamma(k+1) = k!
\end{equation}

\noindent between integers as
\begin{equation}
\Gamma(k+1) = \int\limits_{0}^{\infty} e^{-t} t^{k} dt
\end{equation}

\subsection{Bayesian Information Criterion (BIC)}

The Bayesian Information Criterion\cite{BIC} (BIC), and its close cousin the Akiake Information Criterion\cite{AIC} (AIC), can both be used to evaluate competing models on the same measurements. In this application, we find that the BIC is more effective than the AIC. 

The BIC and AIC penalize larger numbers of parameters fitting the same number of measurements, and thus prevent over-fitting. They become necessary only when the  competing models have differing numbers of parameters, precluding a simple likelihood comparison. A caveat is that since they measure relative performance, not absolute performance, they are unable to warn if all the models fit poorly. 

The BIC equation employed is:

\begin{equation} 
\label{eqnBIC1}
    \mathrm{BIC} = -2\ln(p_z|\theta)+
          n_\theta \ln(n_z)-
          n_\theta \ln(2 \pi)
\end{equation}

\noindent where $p\left(\theta|z\right)$ is evaluated at the ML estimate of $\theta$, and $n_\theta$ and $n_z$ are the number of parameters and measurements respectively.

The BIC is used exclusively in our work to determine the number of observed recoils a neutron experiences within the mTC, and to assist in distinguishing gammas from neutrons as will be explained in the MC results section. Even though we are only interested in the first two scatters, we consider it a prerequisite to first identify all the scatters before isolating the first two from the rest.

\subsection{Candidate Efficiency}

Not all neutrons entering the mTC are viable candidates for estimation. Due to the small size of the mTC, many neutrons leave the detector volume before capture, thus depriving the observer of the characteristic delayed bookend which can assist in positive neutron identification. While not specifically required for neutron direction and energy estimates, an observed capture can identify a particle as a neutron with greater confidence than relying solely on the scatter kinematics. 

A neutron may stay within the mTC volume yet not capture within the available recording time (12 $\mu$s for the mTC). In other cases, a neutron may be identified correctly, yet may simply not deposit adequate energy for reasonable estimation. To exclude these low-quality neutrons from our measurement pool, certain `candidate criteria' must be established.

As a result of these criteria, only about 6\% of 1 MeV neutrons become viable candidates in the mTC, significantly reducing the available measurement population. Table \ref{table:candidate} includes a complete list of candidate criteria applied to our simulated neutrons and the resultant candidate efficiencies (the fraction that pass).

\begin{table} [!htbp]
\fontsize{10}{10}\selectfont
\begin{center}
\begin{tabular}{l|c|c|c|c|c}
\bf Criteria 				& \bf Bk  	& \bf Pu    	& \bf 1 MeV & \bf 3 MeV  	& \bf 5 MeV  \\
\hline
$>20$ PE					& .19		& .38 	    	& .43		& .33 	    	& .27		\\
$>10 P_1$ PE				& .16		& .33 	    	& .36		& .30 	    	& .25		\\
$>5 P_2$ PE					& .19		& .38 	    	& .43		& .32 	    	& .27		\\
$t_2-t_1>1$ns				& .14		& .31 	    	& .38		& .22 	    	& .15		\\
$P_2-P_1>10$mm				& .18		& .36 	    	& .42		& .30 	    	& .24		\\
$dE_0^1>100$keV				& .14		& .28 	    	& .31		& .27 	    	& .23		\\
$dE_0^1/dE_1^2 > .20 $		& .15		& .28 	    	& .32		& .22 	    	& .19		\\
$dE_0^1/E_0 < .9 $			& .18		& .36 	    	& .40		& .32 	    	& .26		\\
$dE_0^1/E_0 > .1 $			& .12		& .26 	    	& .33		& .19 	    	& .13		\\
\hline 
\bf Combined				& \bf .07	& \bf .16  		& \bf .22 	& \bf .12   	& \bf .07  	\\
\end{tabular}
\caption{mTC neutron candidate criteria, including time, distance, energy and photon thresholds, and energy ratio thresholds. Statistics shown for 1, 3, and 5 MeV neutrons as well as Pu and background (Bk) spectra. Background spectra is composed of cosmic-ray induced neutrons at sea level \cite{Goldhagen2004}.}
\label{table:candidate}
\end{center}
\end{table}

\begin{figure}[htbp!]
\includegraphics[width=\linewidth]{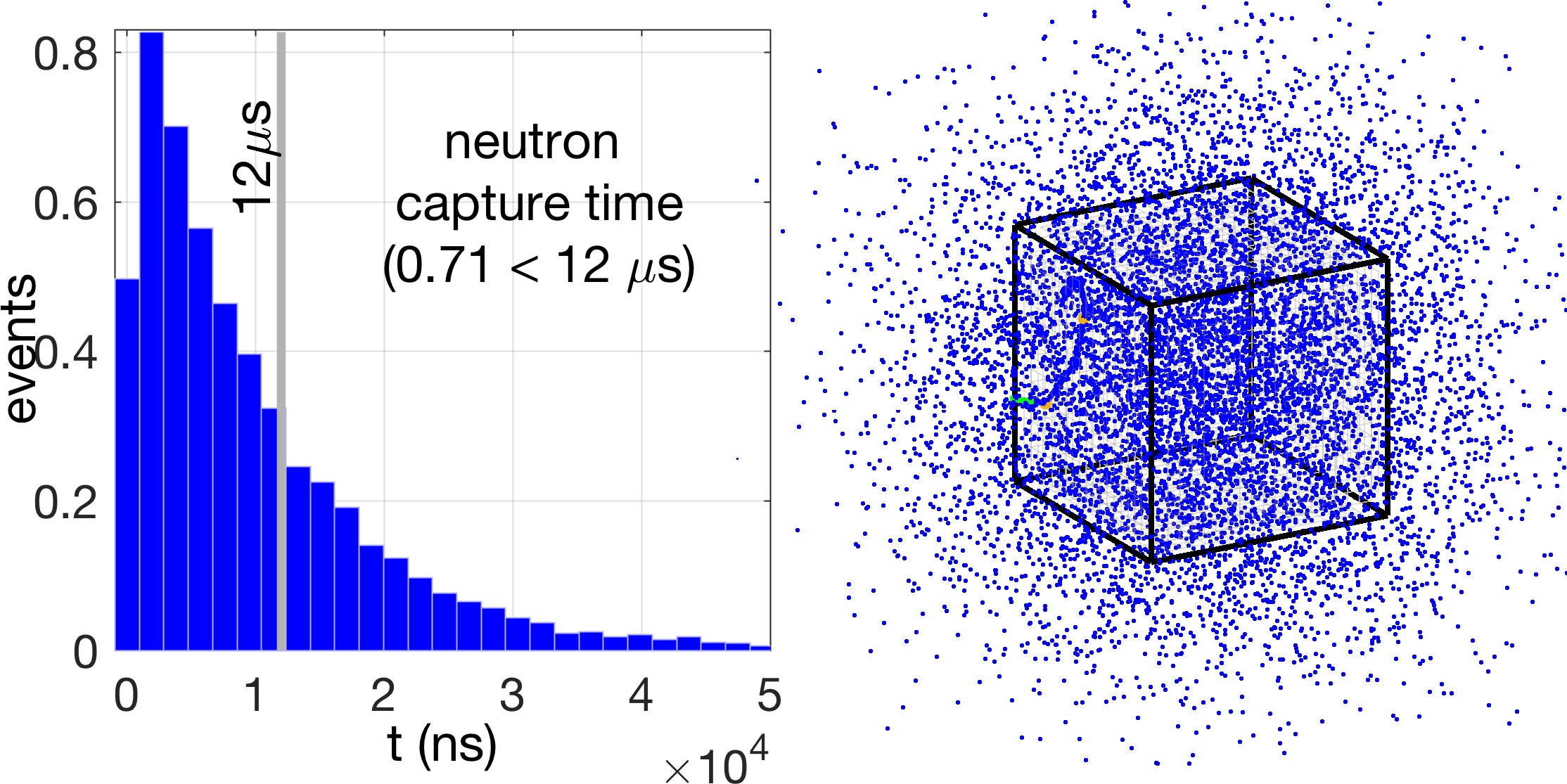}
\caption{Simulated mTC 1 MeV neutron capture times ({\it left}) and locations ({\it right}). Note the small size of mTC allows many neutrons to escape the detector volume.}
\label{candidate21}
\end{figure}

Figure \ref{candidate21} shows 1 MeV neutron capture times and locations in the 1\% EJ-254 mTC. The mTC electronics are limited to 12 $\mu$s of recording time. Note the number of neutrons captured past the mTC time and space limits. Only 70\% of neutrons are captured within $12\mu$s in a 1\% EJ-254 mTC, and only 12\% of neutrons stay inside, and capture within the mTC.

\subsection{MC Results}

\begin{figure}[htbp!]
\includegraphics[width=\linewidth]{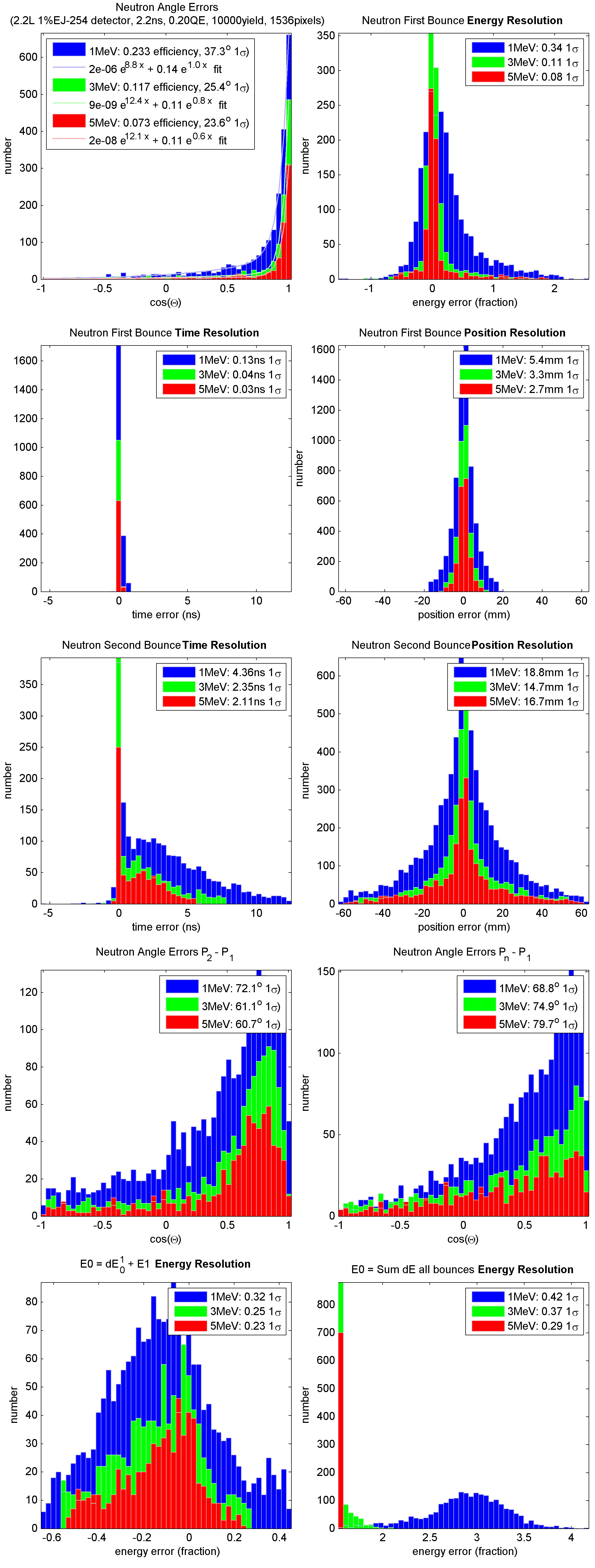}
\caption{mTC MC results.}
\label{mTCMC}
\end{figure}

We use a Monte Carlo simulation of the mTC built with GEANT4\cite{GEANT}, MATLAB\cite{MATLAB} and Python to determine the observability of each of the 5 required reconstruction parameters. Our results are reported as $1\sigma$ resolutions in Table \ref{table:table1}, with the underlying distributions shown in Figure \ref{mTCMC}.

\begin{table} [!htbp]
\fontsize{10}{10}\selectfont
\begin{center}
\begin{tabular}{l|c|c|c|c|c}
\bf $E_0$ 	&\bf $P_1$ 	&\bf $P_2$ 	&\bf $t_1$	&\bf $t_2$ 	&\bf $dE_0^1$	\\
\bf         &\bf (mm) 	&\bf (mm) 	&\bf (ns)  	&\bf (ns)  	&\bf (fraction) \\
\hline 
\bf 1 MeV	& 5.4		& 19 		& 0.13 		& 4.3 		& 0.35 			\\
\bf 3 MeV	& 3.2       & 15 		& 0.04 		& 2.3 		& 0.23 			\\
\bf 5 MeV	& 2.8	    & 16 		& 0.03 		& 2.1 		& 0.19 			\\
\bf Pu		& 2.8	    & 16 		& 0.03 		& 2.1 		& 0.19 			\\
\bf Bk		& 2.8	    & 16 		& 0.03 		& 2.1 		& 0.19 			\\
\end{tabular}
\caption{$P_1$, $P_2$, $t_1$, $t_2$ and $dE_0^1$ mTC MC $1\sigma$ resolutions for 1, 3 and 5 MeV neutrons as well as Pu and background (Bk) spectra. Background spectra is composed of cosmic-ray induced neutrons at sea level \cite{Goldhagen2004}.}
\label{table:table1}
\end{center}
\end{table}

\begin{table} [!htbp]
\fontsize{10}{10}\selectfont
\begin{center}
\begin{tabular}{l|c|c|c|c}
\bf $E_0$ 	& PEs 	& efficiency 	&\bf $\Theta$ 	&\bf $E_0$  \\
\bf         & 		&\bf (fraction) &\bf (deg) 		&\bf (fraction)  \\
\hline
\bf 1 MeV	& 76	& 0.23			& 37 			& 0.32  \\
\bf 3 MeV	& 272	& 0.12     		& 26 			& 0.26  \\
\bf 5 MeV	& 430   & 0.08     		& 24 			& 0.24  \\
\bf Pu		& 170   & 0.12     		& 26 			& 0.26  \\
\bf Bk		& -		& 0.08     		& 24 			& 0.24  \\
\end{tabular}
\caption{$\Theta$ and $E_0$ mTC MC $1\sigma$ resolutions and efficiencies for 1, 3 and 5 MeV neutrons as well as Pu and background (Bk) spectra. Background spectra is composed of cosmic-ray induced neutrons at sea level \cite{Goldhagen2004}.}
\label{table:table2}
\end{center}
\end{table}


Table \ref{table:table2} shows $\Theta$ and $E_0$ resolutions, along with neutron candidate efficiency over the energies tested, and the number PEs captured in the mTC per neutron event.

\begin{figure}[!ht]
\includegraphics[width=\linewidth]{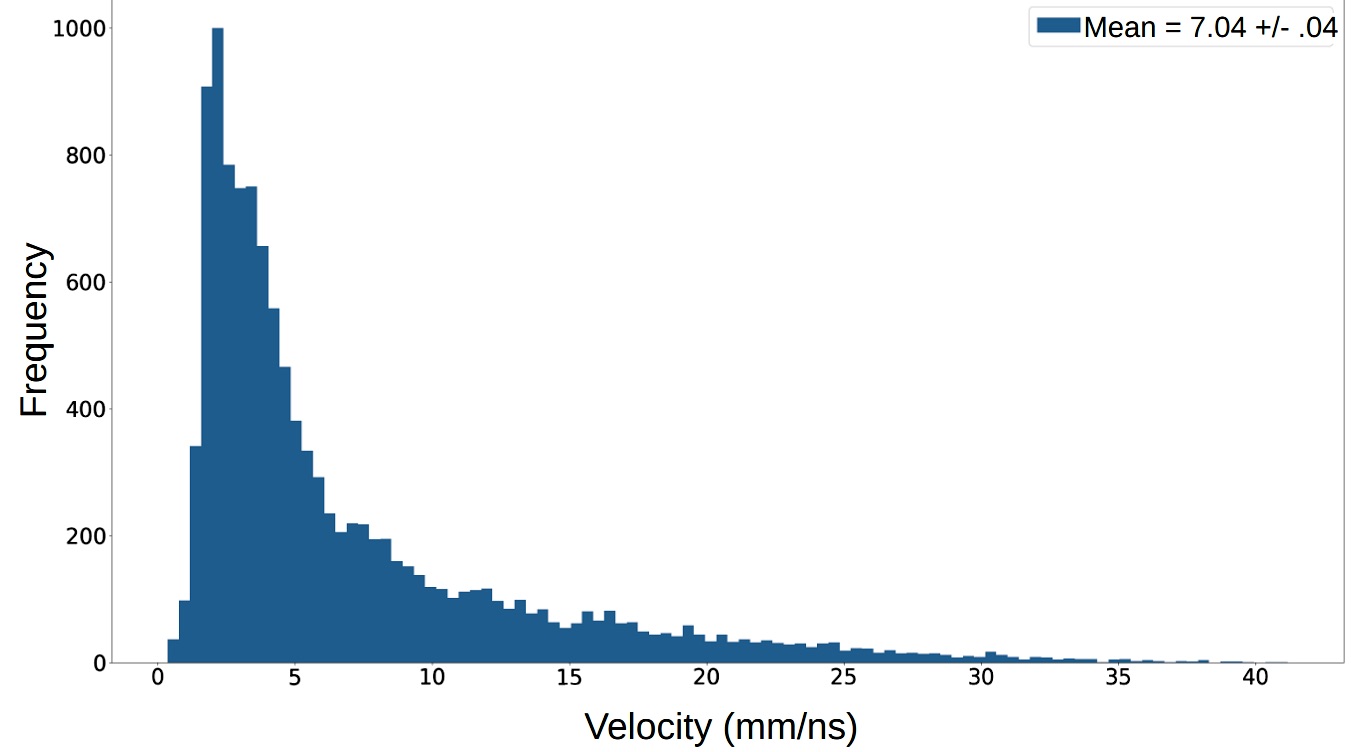}
\caption{A histogram of the simulated neutron velocities between the first and second recoils. The data here allows us to make cuts in our fitting that will eliminate unreasonable velocities in our reconstruction.}
\label{fig:velocity}
\end{figure}

Figure \ref{fig:velocity} shows a histogram of the simulated velocities of the neutron between their first two scatters, and gives us a simple cut with respect to what is a reasonable velocity to expect. We use this to set a velocity rejection cut at $\leq 40$ mm/ns, (note that the speed of light in the scintillator is approximately $190$~mm/ns).  This is gives us an extra buffer against accidental fits for double gamma recoils.

A simulation of 50000 gammas with energies of 0-3 MeV and incident upon the top face of the cube yielded only 3 successful fits using the neutron scatter fitter.  Those 3 events were then promptly rejected using our secondary criteria, that their velocities were on the order of the speed of light in the medium. 

\begin{figure}[htbp!]
\includegraphics[width=\linewidth]{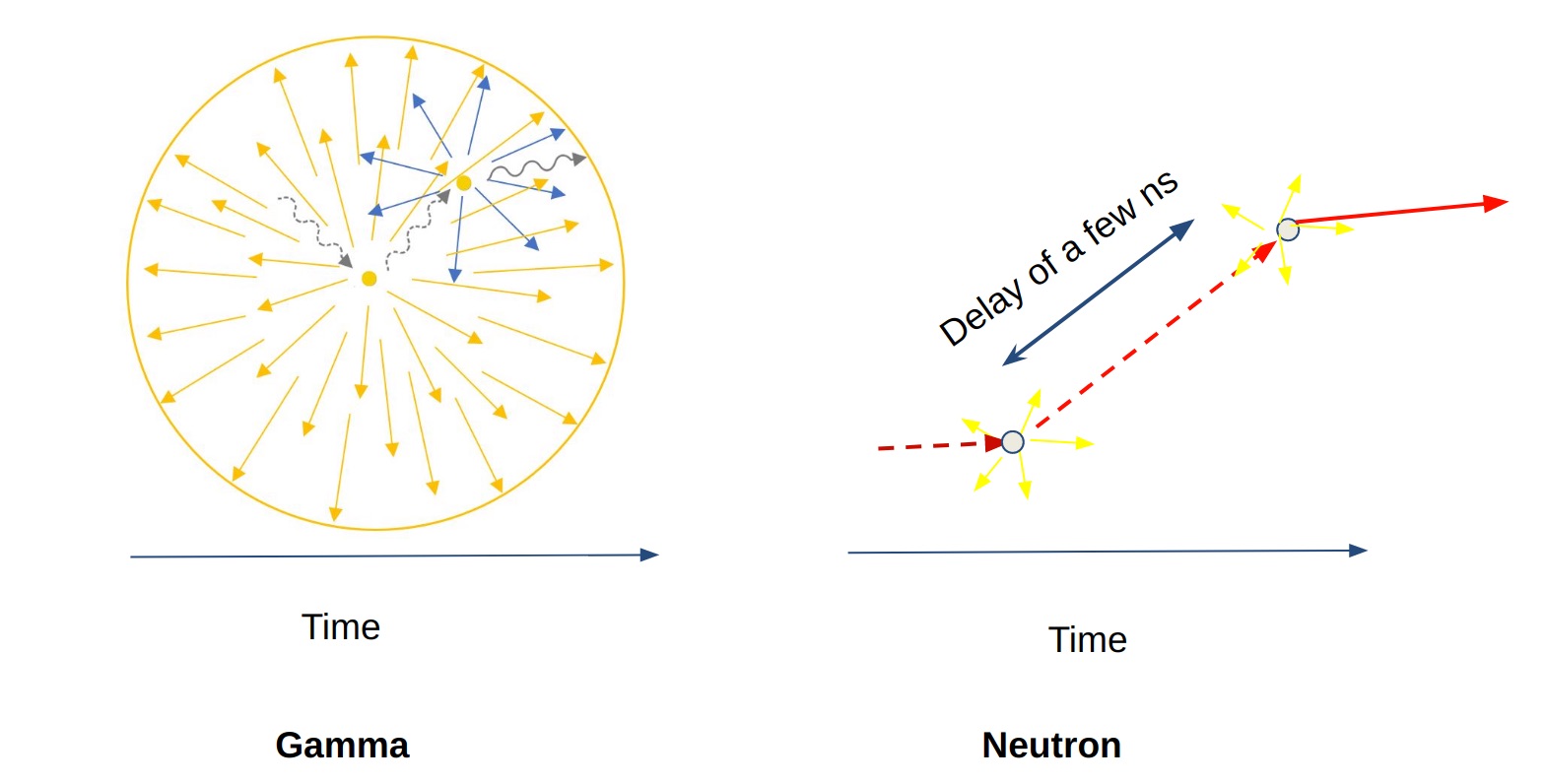}
\caption{Comparison of Gamma and Neutron signatures in plastic scintillator.}
\label{scintgamma}
\end{figure}

A qualitative explanation for why gammas are rejected in our model is given in Figure \ref{scintgamma}. The example gamma Compton scatters twice creating two outgoing scintillation spheres. As light takes approximately a nanosecond to traverse the mTC volume, the scatters are separated in time by only a fraction of a nanosecond. The gamma itself is traveling faster than the photon sphere is expanding and our model will likely fit for a single recoil at a weighted average between the two scintillation points. For the neutron recoils the average time between scatters is approximately 2 nanoseconds. The slower moving neutron allows for the early light from the first recoil to reach the PMTs before the first light produced from the second recoil. As shown in figure \ref{measuredPhotons}, the light arriving from both neutron recoils will always be mixed due to the scintillator decay time.

\begin{figure}[!ht]
    \includegraphics[width=\linewidth]{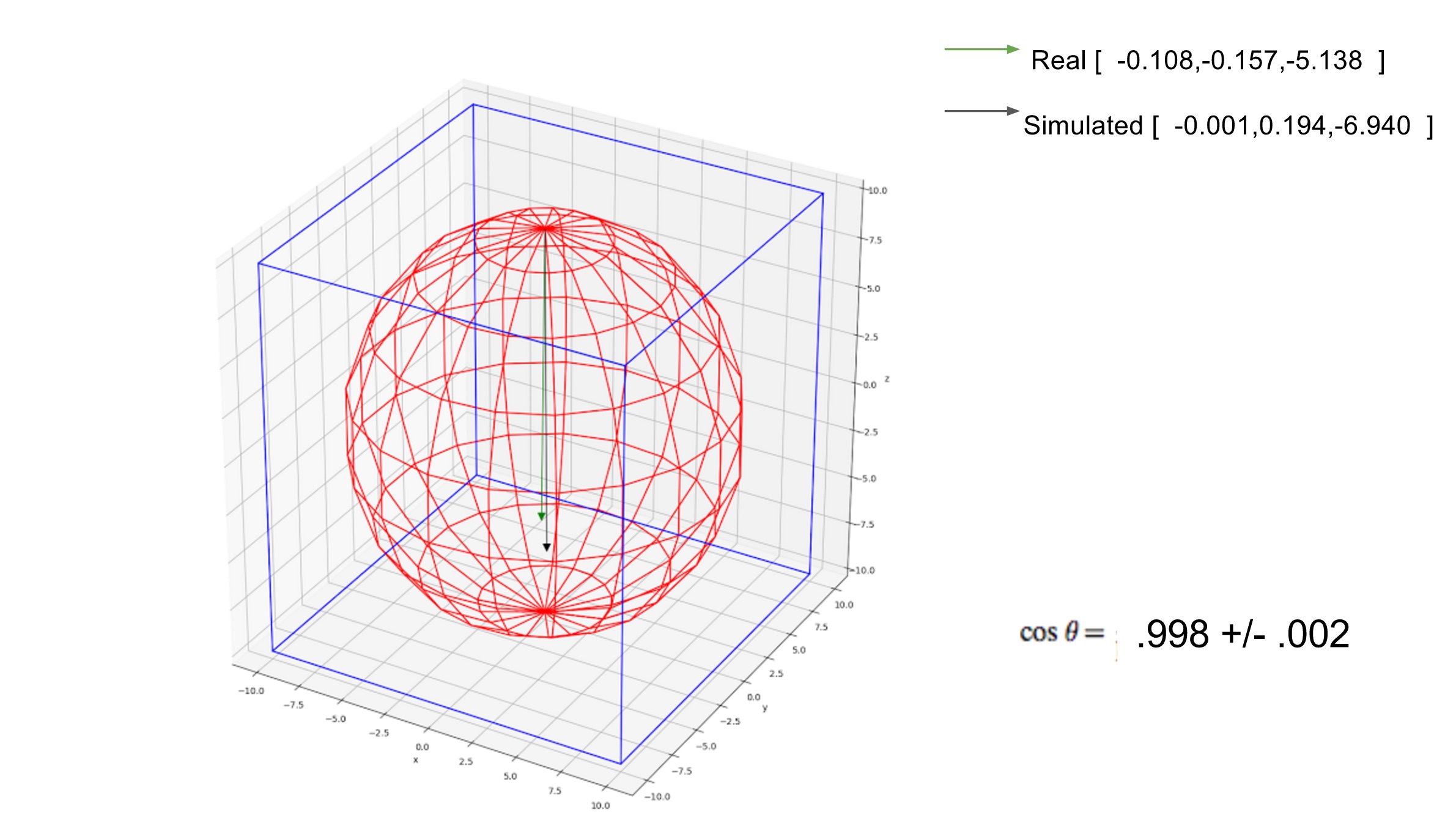}
    \caption{The simulated average directional vector (black) is plotted against the fitted average directional vector (green). The vectors are scaled to the average range between the first two recoils. Clearly there is excellent agreement between the two vectors. $\cos(\theta)$ is .998 +/-.002 corresponding to an angle difference of (-.1,.13) radians or (-5.7,7.4) degrees at a 99.96 percent confidence limit. }
    \label{fig:pointing}
\end{figure}

The fact that we can have confidence that our model only fits for neutrons is an excellent result and allows us to make bold  determinations about neutron directionality. Another avenue for determining neutron directionality is to only consider the average directional vector that the set of vectors $(P_2 - P_1)$. In Figure \ref{fig:pointing} we see a comparison of the fit vs truth for the average directional vector of a set of 10093 neutron events and the agreement is encouraging. $\cos(\theta)$ is .998 +/-.002 corresponding to an angle difference of (-.1,.13) radians or (-5.7,7.4) degrees at a 99.96 percent confidence limit.

\begin{figure}[htbp!]
\includegraphics[width=\linewidth]{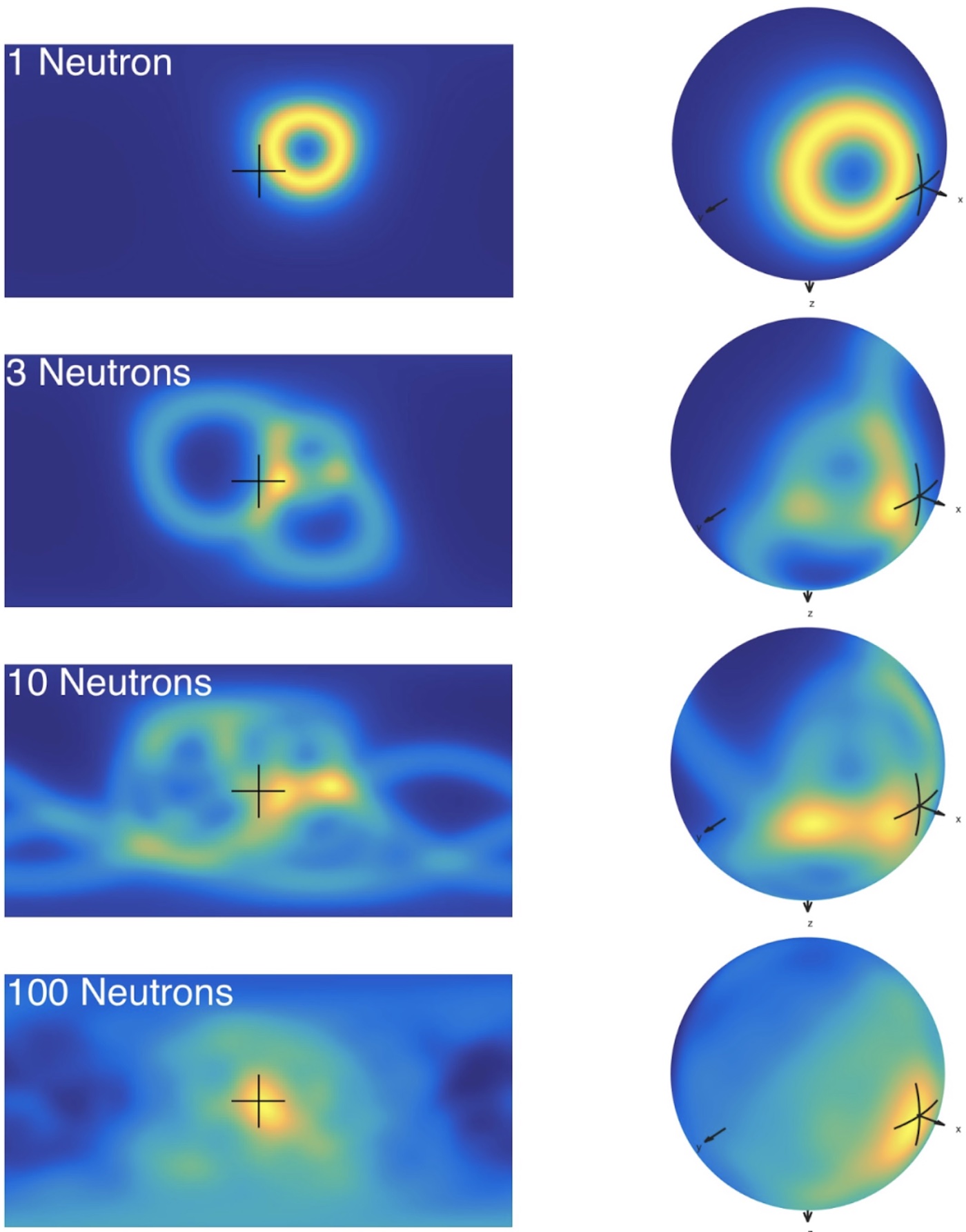}
\caption{Neutron angle-cones projected onto `skymaps' about the detector. One neutron only provides for a cone of possible incoming directions as shown here, with further neutrons serving to reduce this cone to a single direction vector. Black cross represents true incoming direction.}
\label{skymaps}
\end{figure}

Figure \ref{skymaps} shows composite 4pi PDFs over multiple neutron reconstructions. The PDFs for multiple neutrons are simply the summation of individual neutron PDFs. These results include only  `signal' neutrons from a point source of interest. A more comprehensive study would include background neutrons from the Goldhagen\cite{Goldhagen2004} sea level spectrum, as well as background gammas, which are generally present at two orders of magnitude above the neutron background. The addition of both of these backgrounds to the simulation would provide a more accurate picture of detector performance under non-ideal conditions.

\section{Results of Neutron Imaging Experiments}
 
During the neutrino data-collection campaign at NIST the mTC experienced an overheating incident, which resulted in the loss of more than half of the PMTs. The extent of the structural damage to the remaining PMTs (and other components) was unknown. The simple act of turning on the system and performing simple start-up calibrations was nontrivial.  The mTC was shipped back to the University of Hawaii at Manoa for diagnosis. Secondary effects were observed including the increased trigger rate and higher noise. Neutron Imaging Tests were conducted in July 2017 using a californium source placed on various positions around the cube and the results were inconclusive.
 
\begin{figure}[htbp!]
\includegraphics[width=\linewidth]{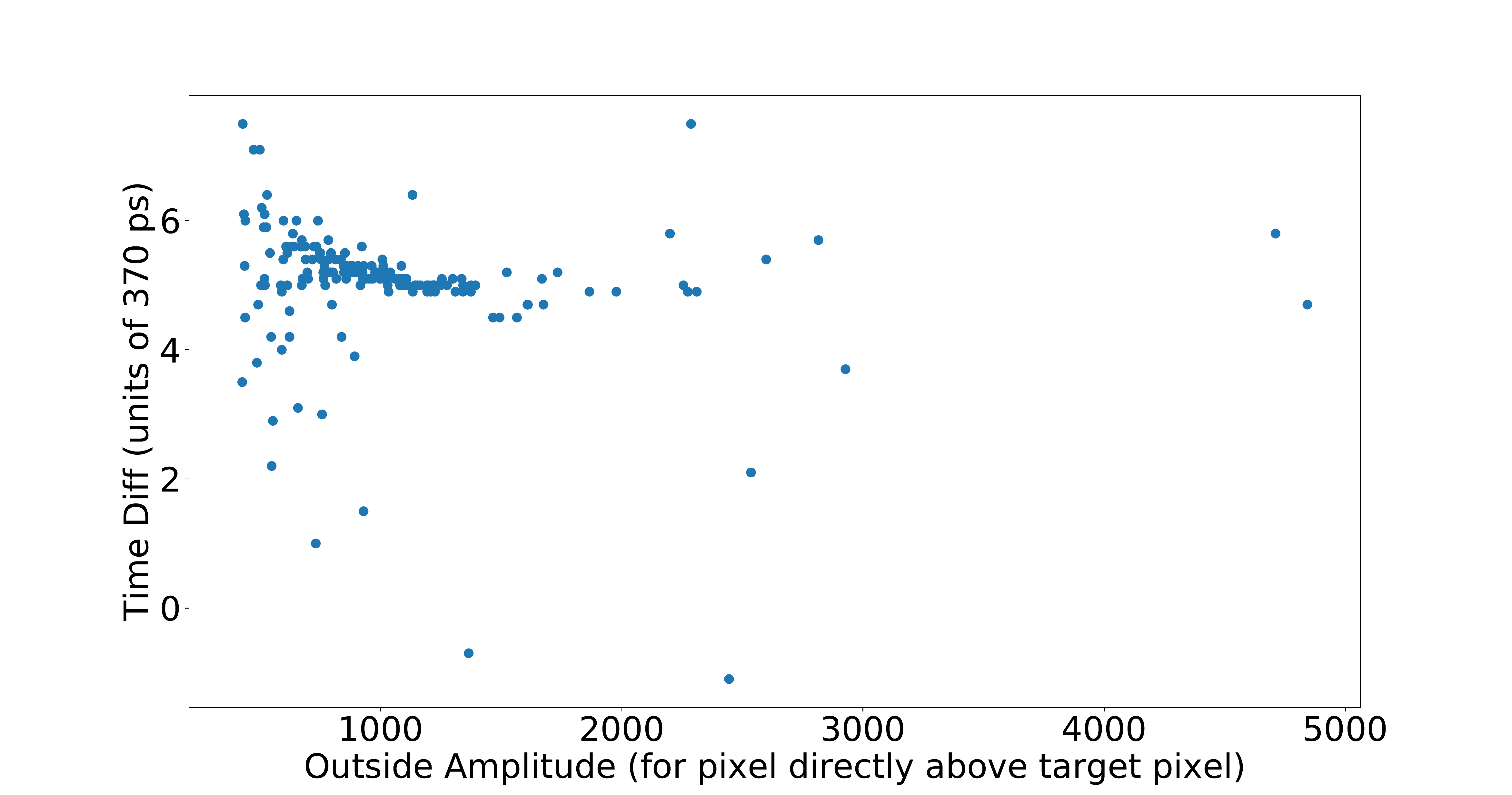}
\caption{Comparison between the relative timing between the target pulse (the pixel in which we inject laser) and the pulse generated in an adjacent pixel, vs. the amplitude of the adjacent pixel. The average delay was 1.76~ns with a standard deviation of 0.70~ns.}
\label{timing}
\end{figure}

Tests were preformed on a surviving PMT to analyze its performance. Laser pulses were injected directly into single MCP channels and a large amount of cross-talk was observed in the surrounding pixels. Figure \ref{timing} quantifies the timing differences between the target pulse (the pixel in which we inject laser) and the undesired pulses generated in the adjacent pixels. The average delay was $1.76$~ns with a standard deviation of .70~ns is a serious issue for our neutron imaging purposes. Since the average time between the low-light generating first and second recoils is on the order of 2~ns, our ability to track neutrons is not possible with the cross-talk response we observed in our PMTs (unknown to us at the time of detector construction). 

\section{Conclusion}
\label{Conclusion}

We presented an efficient model for reconstructing neutron recoils in the mTC which could be adapted for any related design using a single solid volume scintillating target. Our simulations showed that in reconstructing the average directional vector for neutrons entering our scintillating target, the angular agreement between the truth and fit was $\cos(\theta)$ is .998 +/-.002 corresponding to an angle difference of (-.1,.13) radians or (-5.7,7.4) degrees at a 99.96 percent confidence limit. 

Unfortunately, before we undertook neutron imaging tests with a source, our detector suffered a catastrophic overheating event that caused it to behave unreliably. Additionally, the PMTs we used did not behave as expected in terms of their timing response. Nonetheless, we are confident that our model is accurate and efficient for neutron recoil reconstruction.

\subsection{Future detectors}
The miniTimeCube experiment led to the realization that the single scintillation volume might more ideally be replaced by an array of segmented scintillation volumes to allow for more accurate isolation and reconstruction of individual neutron recoils in future  double-scatter detectors. Some of these detector concepts are currently under exploration at the University of Hawaii and various U.S. national labs.

\section{Acknowledgments}
We would like to thank the University of Hawaii and Ultralytics LLC for their support in this effort. Lawrence Livermore National Laboratory is operated by Lawrence Livermore National Security, LLC, for the U.S. Department of Energy, National Nuclear Security Administration under Contract DE-AC52-07NA27344. LLNL-JRNL-762419.

\section{Bibliography}
\bibliography{}

\begin{thebibliography}{0}%
\makeatletter
\providecommand \@ifxundefined [1]{%
 \@ifx{#1\undefined}
}%
\providecommand \@ifnum [1]{%
 \ifnum #1\expandafter \@firstoftwo
 \else \expandafter \@secondoftwo
 \fi
}%
\providecommand \@ifx [1]{%
 \ifx #1\expandafter \@firstoftwo
 \else \expandafter \@secondoftwo
 \fi
}%
\providecommand \natexlab [1]{#1}%
\providecommand \enquote  [1]{``#1''}%
\providecommand \bibnamefont  [1]{#1}%
\providecommand \bibfnamefont [1]{#1}%
\providecommand \citenamefont [1]{#1}%
\providecommand \href@noop [0]{\@secondoftwo}%
\providecommand \href [0]{\begingroup \@sanitize@url \@href}%
\providecommand \@href[1]{\@@startlink{#1}\@@href}%
\providecommand \@@href[1]{\endgroup#1\@@endlink}%
\providecommand \@sanitize@url [0]{\catcode `\\12\catcode `\$12\catcode
  `\&12\catcode `\#12\catcode `\^12\catcode `\_12\catcode `\%12\relax}%
\providecommand \@@startlink[1]{}%
\providecommand \@@endlink[0]{}%
\providecommand \url  [0]{\begingroup\@sanitize@url \@url }%
\providecommand \@url [1]{\endgroup\@href {#1}{\urlprefix }}%
\providecommand \urlprefix  [0]{URL }%
\providecommand \Eprint [0]{\href }%
\providecommand \doibase [0]{http://dx.doi.org/}%
\providecommand \selectlanguage [0]{\@gobble}%
\providecommand \bibinfo  [0]{\@secondoftwo}%
\providecommand \bibfield  [0]{\@secondoftwo}%
\providecommand \translation [1]{[#1]}%
\providecommand \BibitemOpen [0]{}%
\providecommand \bibitemStop [0]{}%
\providecommand \bibitemNoStop [0]{.\EOS\space}%
\providecommand \EOS [0]{\spacefactor3000\relax}%
\providecommand \BibitemShut  [1]{\csname bibitem#1\endcsname}%
\let\auto@bib@innerbib\@empty
\end{thebibliography}%


\begin{thebibliography}{00}

\bibitem{jgl09}  J.G.Learned, High Energy Neutrino Physics with Liquid Scintillation Detectors; arXiv:0902.4009, J.Kumar, J.G.Learned, S. Smith, Phys. Rev. D80 (2009) 113002; J. Peltoniemi arXiv0909.4974 and arXiv0911.4876; M. Wurm, {\it et al}, Acta Phys. Polon. B431 (2010) 1749-1764.

\bibitem{ber13} B. Adams, M. Chollet, A. Elagin, E. Oberla, A. Vostrikov, et al. Review of Scientific Instruments (2013) 84 (6)

\bibitem{dchoo}Double Chooz Proposal, hep-ex/0606025

\bibitem{reno}RENO proposal, arxiv:1003.1391

\bibitem{daya}Daya Bay Collaboration (proposal), arXiv:hep-ex/0701029

\bibitem{kli94}Yu. V. Klimov, et al., Atomic Energy 76 (1994) 123

\bibitem{ber02}A. Bernstein, et al. J. Appl. Phys. 91 (2002) 4672

\bibitem{bow07}N.S. Bowden, et al. , Nucl. Instr. and Meth. A 572 (2007) 985-998

\bibitem{men11}G. Mention, et al. , Phys. Rev. D (2011) 83, 073006

\bibitem{van07}A. Vanier, L. Forman, I. Dioszegi, C. Salwen, and V.J. Ghosh. Calibration and Testing of a Large Area Fast-Neutron Directional Detector. NSS Conference Record, 2007. IEEE

\bibitem{pre85} A.M. Preszler, W.A. Millard, and S.E. Walker, in Proceeedings of the Eleventhg Symposium on Fusion Energy Nov., 1985, 3C04

\bibitem{wal96} S.E. Walker, A.M. Preszler, and W.A. Millard, Double scatter neutron time-of-flight spectrometer as a plasma diagnostic, Rev. Sci.. Instrum, 1986, 57, 1740

\bibitem{mas09} N. Mascarenhas, J. Brennan, K. Krenz, P. Marleau, S. Mrowka. Results With the Neutron Scatter Camera. Nuclear Science, IEEE Transactions, 2009, 56, 3

\bibitem{bla11} M. A. Blackston, B. L. Brown, E. Brubaker, P. A. Hausladen, P. Marleau, J. Newby, “Fast-Neutron Coded-Aperture Imaging for Warhead Counting”, INNM Conference Record, 2011, Desert Springs, CA

\bibitem{bru09} E. Brubaker, P. Marleau, et al., “Calibration and Simulation of a Coded Aperture Neutron Imaging System”, Nuclear Science Symposium Conference Record, 2009. IEEE, Orlando, FL

\bibitem{gao11} Report to Congressional Requesters, Technology Assesment, Neutron Detectors Alternatives to using helium-3. 2011

\bibitem{kou09} R. Kouzes, J. Ely, A. Lintereur, and D. Stephens 2009. Neutron Detector Gamma Insensitivity Criteria. PNNL-18903. Richland, Wash: Pacific Northwest National Laboratory.

\bibitem{she10} D. Shea and D. Morgan. The Helium-3 Shortage: Supply, Demand, and Options for Congress. 2010, Congressional Research Service.  

\bibitem{smi05} T. Smith, et al., Nucl. Instr. and Meth. A 550, 2005, 90

\bibitem{bow10} N.S. Bowden, M. Heffner, G. Carosi, D. Carter, P. O'Malley, J. Mintz, M. Foxe, and I. Jovanovic. Directional fast neutron detection using a time projection chamber. Nucl. Instr. and Meth. A 624, 2010, 153

\bibitem{hun08} S. Hunter, G. de Nolf, L. Barbier, J. Link, S. Floyd, N. Guardala, M. Skopec, and B. Stark. Neutron Imaging Camera, Proc. SPIE 6954 (2008) 695415

\bibitem{BIC} Schwarz, Gideon E. (1978). ``Estimating the dimension of a model". Annals of Statistics 6 (2): 461–464. doi:10.1214/aos/1176344136. MR 468014.

\bibitem{AIC} Akaike, H., 1977. ``On entropy maximization principle". In: Krishnaiah, P.R. (Editor). Applications of Statistics, North-Holland, Amsterdam, pp. 27–41.

\bibitem{GEANT} GEANT4 - A Simulation Toolkit, S. Agostinelli \textit{et al.}, Nuclear Instruments and Methods A 506 (2003) 250-303 \url{http://geant4.cern.ch/}

\bibitem{MATLAB} MATLAB, a computer program \url{http://www.mathworks.com}

\bibitem{EJ-254} Eljen Technology. EJ-254 Boron Loaded Plastic Scintillator,  \url{http://www.eljentechnology.com/}


\bibitem{Bar-Shalom} ``Estimation with Applications to Tracking and Navigation", Bar-Shalom, Li, Kirubarajan  (2001) ISBN-10: 047141655X 

\bibitem{Goldhagen2004} Gordon, M.S.; Goldhagen, P.; Rodbell, K.P.; Zabel, T.H.; Tang, H. H K; Clem, J. M.; Bailey, P., ``Measurement of the flux and energy spectrum of cosmic-ray induced neutrons on the ground," Nuclear Science, IEEE Transactions on , vol.51, no.6, pp.3427,3434, Dec. 2004, doi: 10.1109/TNS.2004.839134,  \url{http://ieeexplore.ieee.org/stamp/stamp.jsp?tp=&arnumber=1369506&isnumber=29973}



\bibitem{jocher2013}
Glenn R. Jocher, Daniel A. Bondy, Brian M. Dobbs, Stephen T. Dye, James A. Georges III, John G. Learned, Christopher L. Mulliss, Shawn Usman, Theoretical antineutrino detection, direction and ranging at long distances, Physics Reports, Volume 527, Issue 3, 20 June 2013, Pages 131-204, ISSN 0370-1573, \url{http://dx.doi.org/10.1016/j.physrep.2013.01.005}.
(\url{http://www.sciencedirect.com/science/article/pii/S0370157313000240})

\bibitem{li2016}
V. A. Li, R. Dorrill, M. J. Duvall, J. Koblanski, S. Negrashov, M. Sakai, S. A. Wipperfurth, K. Engel, G. R. Jocher, J. G. Learned, L. Macchiarulo, S. Matsuno, W. F. McDonough, H. P. Mumm, J. Murillo, K. Nishimura, M. Rosen, S. M. Usman and   G. S. Varner, 
Invited Article: miniTimeCube, Review of Scientific Instruments 87, 021301 (2016), \url{https://doi.org/10.1063/1.4942243}
(\url{http://aip.scitation.org/doi/10.1063/1.4942243})

\bibitem{Askins:2015bmb} 
  M.~Askins {\it et al.} [WATCHMAN Collaboration],
  arXiv:1502.01132 [physics.ins-det].


\bibitem{Barrett:2013zka} 
  M.~Barrett [Belle-II iTOP Group],
  arXiv:1310.4542 [physics.ins-det].

\bibitem{Li:2018dwm} 
  V.~Li,
  ``miniTimeCube: Building The World's Smallest Neutrino Detector,'' PhD thesis, INSPIRE record 1666321.



  
\end{thebibliography}

\end{document}